\documentclass[11pt,a4paper,aps,prd,superscriptaddress,nofootinbib,preprintnumbers,longbibliography]{revtex4-2}
\usepackage{geometry}
\usepackage[utf8]{inputenc}
\usepackage[T1]{fontenc}
\usepackage{verbatim}
\usepackage{amsmath,amssymb,amsfonts}	
\usepackage{graphicx}
\usepackage{color}
\usepackage{xcolor}
\usepackage{physics}
\usepackage{booktabs}

\usepackage{appendix}
\usepackage{slashed}
\usepackage{tikz}
\usetikzlibrary{shapes}
\usepackage[colorlinks=true,citecolor=blue,linkcolor=purple,urlcolor=purple]{hyperref}

\makeatletter
\def\l@subsubsection#1#2{}
\makeatother

\begin{document}
\preprint{FERMILAB-PUB-22-927-T}\preprint{CPPC-2022-13}
\heavyrulewidth=.08em
\lightrulewidth=.05em
\cmidrulewidth=.03em
\belowrulesep=.65ex
\belowbottomsep=0pt
\aboverulesep=.4ex
\abovetopsep=0pt
\cmidrulesep=\doublerulesep
\cmidrulekern=.5em
\defaultaddspace=.5em

\baselineskip 0.7cm

\bigskip

\title{Lepton-flavour-violating tau decays from triality}

\author{Innes Bigaran}
\email[e-mail: ]{ibigaran@fnal.gov}
\affiliation{Northwestern University, Department of Physics \& Astronomy, 2145 Sheridan Road, Evanston, IL 60208, USA}
\affiliation{Theoretical Physics Department, Fermilab, P.O. Box 500, Batavia, IL 60510, USA}
\affiliation{ARC Centre of Excellence for Dark Matter Particle Physics, School of Physics, The University of Melbourne, Victoria 3010, Australia}

\author{Xiao-Gang He}
\email[e-mail: ]{hexg@sjtu.edu.cn}
\affiliation{Tsung-Dao Lee Institute, and School of Physics and Astronomy, Shanghai Jiao Tong University, Shanghai 200240, China}
\affiliation{National Center for Theoretical Sciences, and Department of Physics, National Taiwan University, Taipei 20617, Taiwan}

\author{\mbox{Michael A.~Schmidt}}
\email[e-mail: ]{m.schmidt@unsw.edu.au}
\affiliation{Sydney Consortium for Particle Physics and Cosmology, School of Physics, The University of New South Wales,
Sydney, New South Wales 2052, Australia
}

\author{German Valencia}
\email[e-mail: ]{german.valencia@monash.edu}
\affiliation{School of Physics, Monash University, Wellington Road, Clayton, Victoria 3800, Australia}

\author{Raymond Volkas}
\email[e-mail: ]{raymondv@unimelb.edu.au}
\affiliation{ARC Centre of Excellence for Dark Matter Particle Physics, School of Physics, The University of Melbourne, Victoria 3010, Australia}

\begin{abstract}
Motivated by flavour symmetry models, we construct theories based on a low-energy limit featuring lepton flavour triality that have the flavour-violating decays $\tau^\pm \to \mu^\pm \mu^\pm e^\mp$ and $\tau^\pm \to e^\pm e^\pm \mu^\mp$ as the main phenomenological signatures of physics beyond the standard model. These decay modes are expected to be probed in the near future with increased sensitivity by the Belle II experiment at the SuperKEKB collider. The simple standard model extensions featured have doubly-charged scalars as the mediators of the above decay processes. The phenomenology of these extensions is studied here in detail.

\end{abstract}

\maketitle

\newpage

\tableofcontents
\newpage

\section{Introduction}

The search for charged-lepton flavour violation~(cLFV) is an important component of the general program of seeking signals of physics beyond the standard model (BSM). The discovery of neutrino flavour oscillations established that the family lepton numbers $L_e$, $L_\mu$ and $L_\tau$ are not individually conserved. With the charged and neutral leptons coexisting in the same weak-isospin doublet, we thus expect that these quantities will also not be conserved in the charged-lepton sector. However, for such processes to be experimentally observable, lepton flavour violating~(LFV) physics beyond that responsible for family-lepton number violating neutrino mass generation must exist.

There are stringent existing constraints on electron-muon cLFV processes such as $\mu \to e \gamma$ and $\mu \to eee$. Existing data on cLFV involving $\tau$ leptons are less constraining, but the sensitivity to such processes is expected to increase significantly as the Belle II experiment accumulates more data. The purpose of this paper is to construct and analyse some simple standard model~(SM) extensions that have the decays
\begin{align}
    \tau^\pm \to \mu^\pm \mu^\pm e^\mp,\qquad \tau^\pm \to \mu^\mp e^\pm e^\pm 
    \label{eq:decays}
\end{align}
as the main phenomenological signatures of BSM physics. We summarise the current bounds and projected Belle II sensitivities in Table~\ref{tab:LFVtau}.

\begin{table}[b!]
    \centering
\setlength{\tabcolsep}{10pt}
\renewcommand{\arraystretch}{1.5}
	\begin{tabular}{|c|c|c|}
 \hline
Observable & Present constraint & Projected sensitivity \\
\hline
BR$( \tau^- \to \mu^- \mu^- e^+)$ &$< 1.7 \times 10^{-8}$~\cite{Hayasaka:2010np}& $2.6\times 10^{-10}$~\cite{Banerjee:2022xuw}\\
BR$(\tau^- \to \mu^+ e^-e^- )$ & $< 1.5 \times 10^{-8}$~\cite{Hayasaka:2010np} & $2.3\times 10^{-10}$~\cite{Banerjee:2022xuw}\\
\hline
 \end{tabular}
    \caption{We list the cLFV tau decays of interest, where the projected reach reflects the expected sensitivity with $50$ ab$^{-1}$ data from the Belle II collaboration~\cite{Banerjee:2022xuw} assuming a phase space distribution for the 3-body decay (see discussion in Sec.~\ref{sec:phase space}).}
    \label{tab:LFVtau}
\end{table}

To achieve our aim, a symmetry is needed that permits the above decays but also prevents other processes that are subject to strong constraints. A simple choice for such a symmetry is \emph{lepton flavour triality}. This has, for example, been discussed in the context of $A_4$ flavour models~\cite{Altarelli:2005yx,He:2006dk,Ma:2010gs,deAdelhartToorop:2010jxh,deAdelhartToorop:2010nki,Cao:2011df,Holthausen:2012wz,Pascoli:2016wlt,Muramatsu:2016bda} which are broken to a $Z_3$ subgroup in the charged lepton sector and $Z_2$ in the neutrino sector.

Motivated by an eventual embedding into a more complete flavour symmetry model, we introduce a $Z_3$ symmetry in the lepton sector which distinguishes the three families via their $Z_3$ \emph{triality charges} (denoted $T$). The first (second) [third] generation of leptons has charge $T= 1$ ($2$) [$3$]. These charges correspond to the transformations
\begin{align}\label{eq:triality}
    L \to \omega^T L,\quad e_R \to \omega^T e_R
\end{align}
where $\omega=e^{2\pi i/3}$ is the third root of unity, and the $L$ and $e_R$ are left-handed (LH) lepton doublets and right-handed (RH) charged lepton singlets, 
respectively. Thus all first-family leptons transform via $\omega$, all second-family leptons transform via $\omega^2$, and all third-family leptons are triality singlets, transforming via $\omega^3 = 1$. A discussion of neutrino masses is postponed to Sec.~\ref{sec:neutrino}.
The Higgs doublet, $H$, and all quark fields are also triality singlets.

Given these triality assignments, the leptonic Yukawa terms in the Lagrangian are
\begin{align}
    \mathcal{L} \supset y_{ei} \bar L_i e_{Ri}  H 
    +\mathrm{h.c.}
    ,
\end{align}
where $i,j=1,2,3$ with repeated indices summed, 
and $y_{ei}$ 
denote the charged lepton 
Yukawa couplings.
The $Z_3$ symmetry forces the charged lepton mass matrix to be diagonal, and we thus identify the first (second) [third] generation of charged leptons with the electron (muon) [tau] lepton.

\begin{figure}[t!]
    \centering  \includegraphics[width=0.35\textwidth]{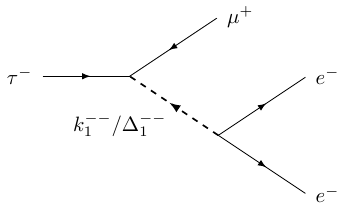}
    \hspace{1cm}
\includegraphics[width=0.4\textwidth]{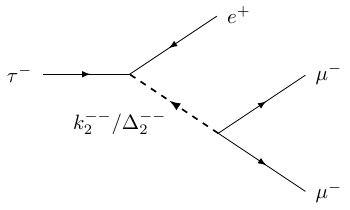}
    \caption{Tree-level contributions to the process $\tau^-\to\mu^+ e^- e^- $ ($\tau^-\to \mu^- \mu^- e^+$) for the $T=1$ ($T=2$) electroweak singlet scalar models, utilising $k_i$, and triplet models, utilising $\Delta_i$. Note that the singlet models couple to RH leptons, whereas the triplets couple to LH leptons.}
    \label{fig:FeynmanDiagrams}
\end{figure}

This simple model outlined above forms the basis for extensions that feature observable decays of the form of Eq.~(\ref{eq:decays}).  The simplest models are based on scalar bileptons~\cite{Cuypers:1996ia}, in particular doubly-charged scalars\footnote{Doubly-charged scalars as mediators of cLFV tau decays have been studied, for example, in Refs.~\cite{Akeroyd:2006bb,Nebot:2007bc,Akeroyd:2009nu,Heeck:2016xwg,Crivellin:2018ahj,BhupalDev:2018vpr}.} as illustrated in Fig.~\ref{fig:FeynmanDiagrams}. 
In Sec.~\ref{sec:singlet} we discuss the simplest realisation based on an electroweak singlet and
in Sec.~\ref{sec:triplet} we present a model based on an electroweak triplet.
The phase space for the cLFV leptonic $\tau$ decays is discussed in Sec.~\ref{sec:phase space} and different possibilities to introduce neutrino masses are introduced in Sec.~\ref{sec:neutrino} and in Sec.~\ref{sec:conclusions} we summarise and conclude. Technical details are reported in the appendices.

\section{Electroweak singlet models}
\label{sec:singlet}

The simplest model that has cLFV $\tau$ decays as the dominant BSM signature features a doubly-charged scalar weak-isospin singlet $k_T$ with lepton triality charge $T$, hypercharge $Y(k_T)=2$ and a mass $m_{k_T}$. Depending on the lepton triality assignment for the doubly-charged scalar $k_T$, there are different cLFV $\tau$ decay modes. Note that we only consider $T=1$ and $T=2$, because the $T=0$ case does not result in cLFV $\tau$ decays.

\subsection{\texorpdfstring{$T=1$}{T=1} singlet model}
\label{subsec:T1Singlet}
For triality $T=1$ the Yukawa couplings of the doubly-charged scalar $k_1$ are
\begin{equation}\label{eq:Lag0}
    \mathcal{L}_{k_1} = \frac12 \left( 2 f_1 \overline{(\tau_{R})^c} \mu_{R} + f_2 \overline{(e_{R})^c} e_{R} \right) k_1 + \mathrm{h.c.} .
\end{equation}
which induce the decays $\tau^\pm \to \mu^\mp e^\pm e^\pm $ via the LH diagram in Fig.~\ref{fig:FeynmanDiagrams}. 
Note that we may, without loss of generality, set the coupling constants $f_{1,2}$ to be real-valued and positive, which we do from now on. One of the phases of $f_{1,2}$ may be absorbed into $k_1$, and the other absorbed into a RH charged lepton. Ultimately, the leptonic CP violation can be taken to reside entirely in the PMNS matrix and the decay modes of the heavy neutral leptons.

For energies below the mass of $k_1$, we can employ standard model effective field theory (SMEFT) and match to low-energy effective field theory (LEFT), the concepts and notation of which are reviewed in Appendices~\ref{sec:SMEFT} and \ref{sec:LEFT}. The LEFT Wilson coefficient for the relevant cLFV decay mode  
\begin{align}
    C^{\rm VRR}_{ee, 1312} =\frac{f_1 f_2}{4m_{k_1}^2}
\end{align}
contributes to the branching ratio\footnote{Neglecting the electron mass, the effect of the muon mass is the same for $\mathrm{BR}(\tau^\pm \to  \mu^\mp e^\pm e^\pm)$ and $\mathrm{BR}(\tau^- \to \mu^- \bar\nu_\mu \nu_\tau)$.}
\begin{align}
\label{eq:taubarmuee_k1}
   \mathrm{BR}(\tau^\pm \to \mu^\mp e^\pm e^\pm ) & =
\frac{f_1^2 f_2^2}{64 G_F^2 m_{k_1}^4} \mathrm{BR}(\tau^- \to \mu^- \bar\nu_\mu \nu_\tau) .
\end{align}

\begin{figure}[t!]
    \centering
    \includegraphics[width=0.48\textwidth]{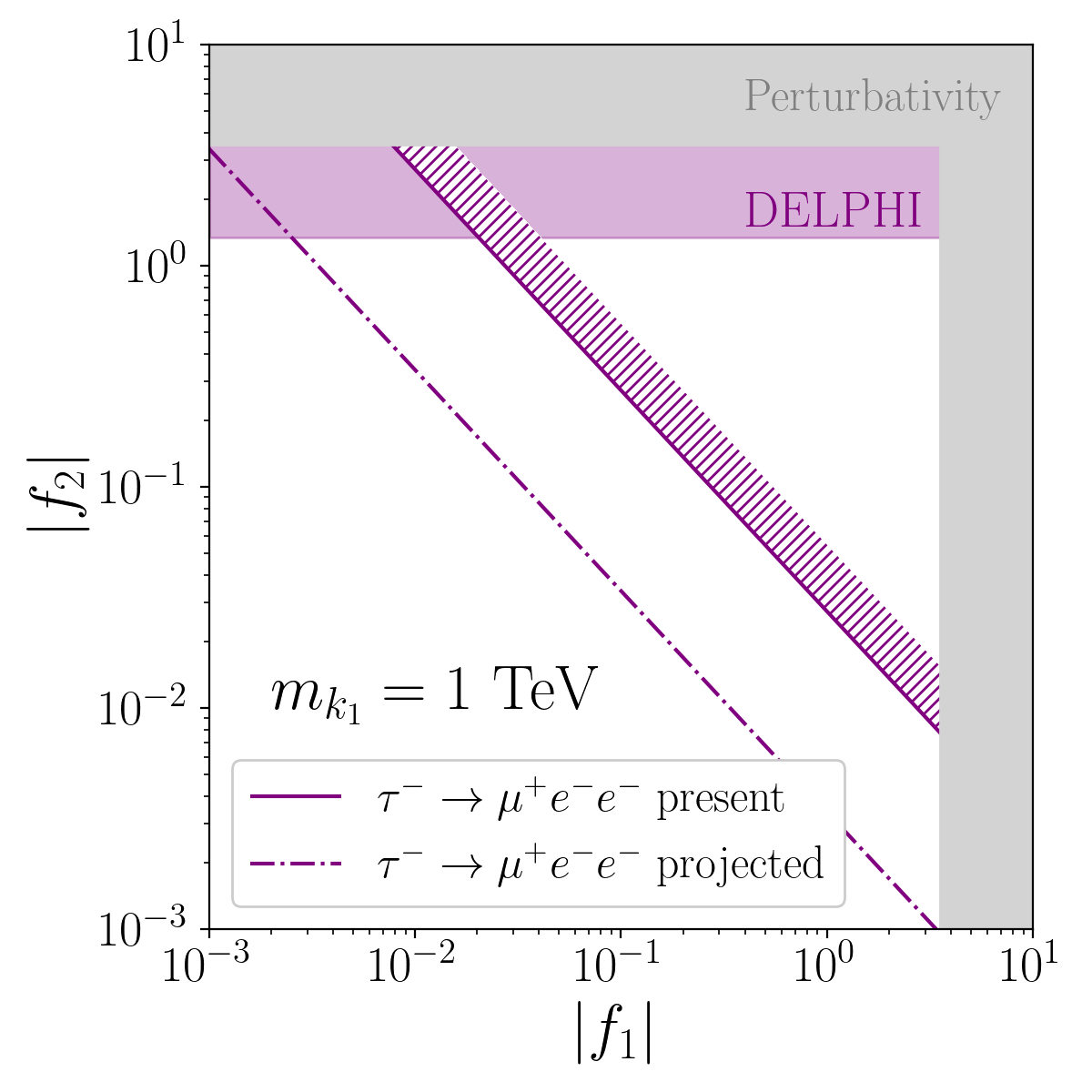}
   \includegraphics[width=0.48\textwidth]{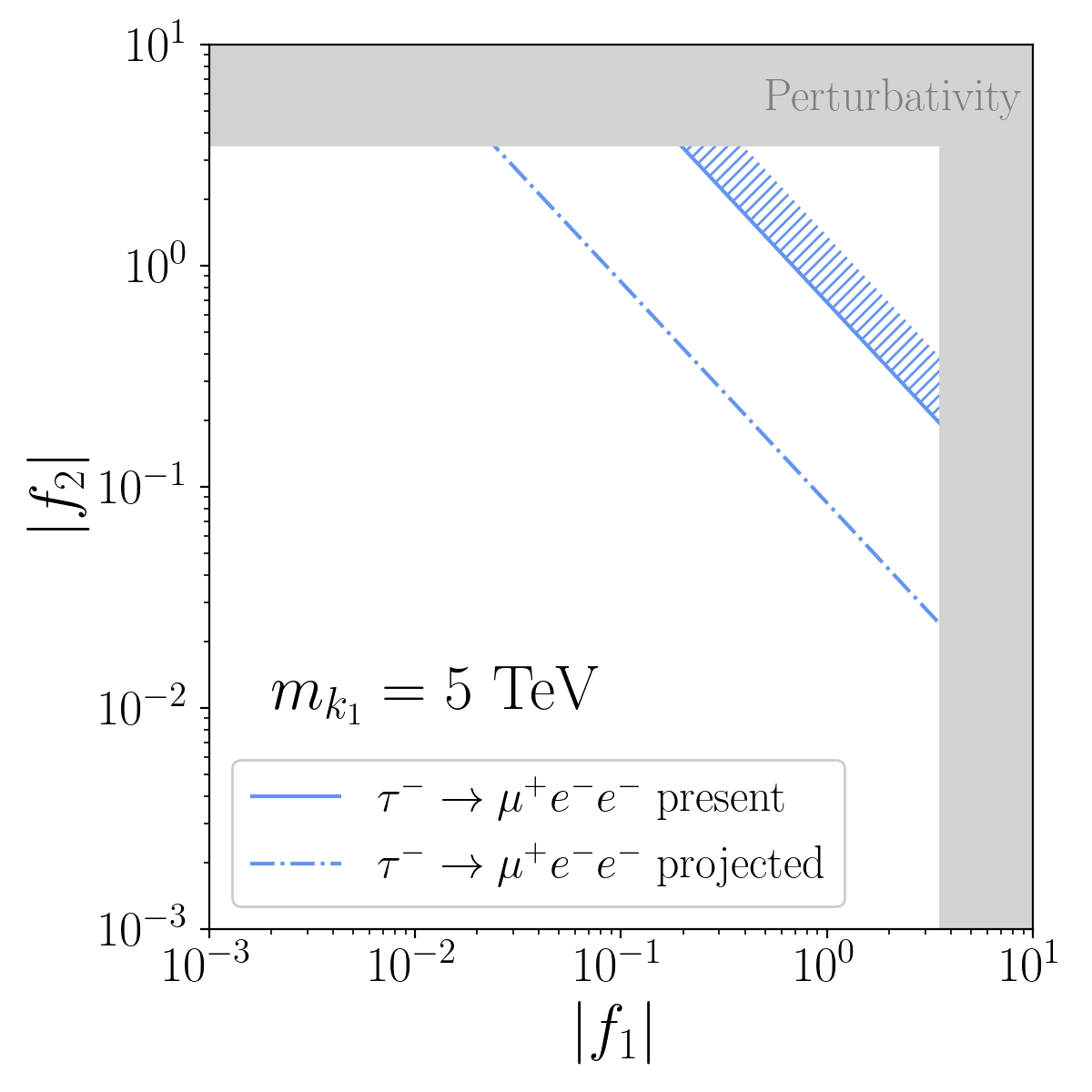}
   \includegraphics[width=0.48\textwidth]{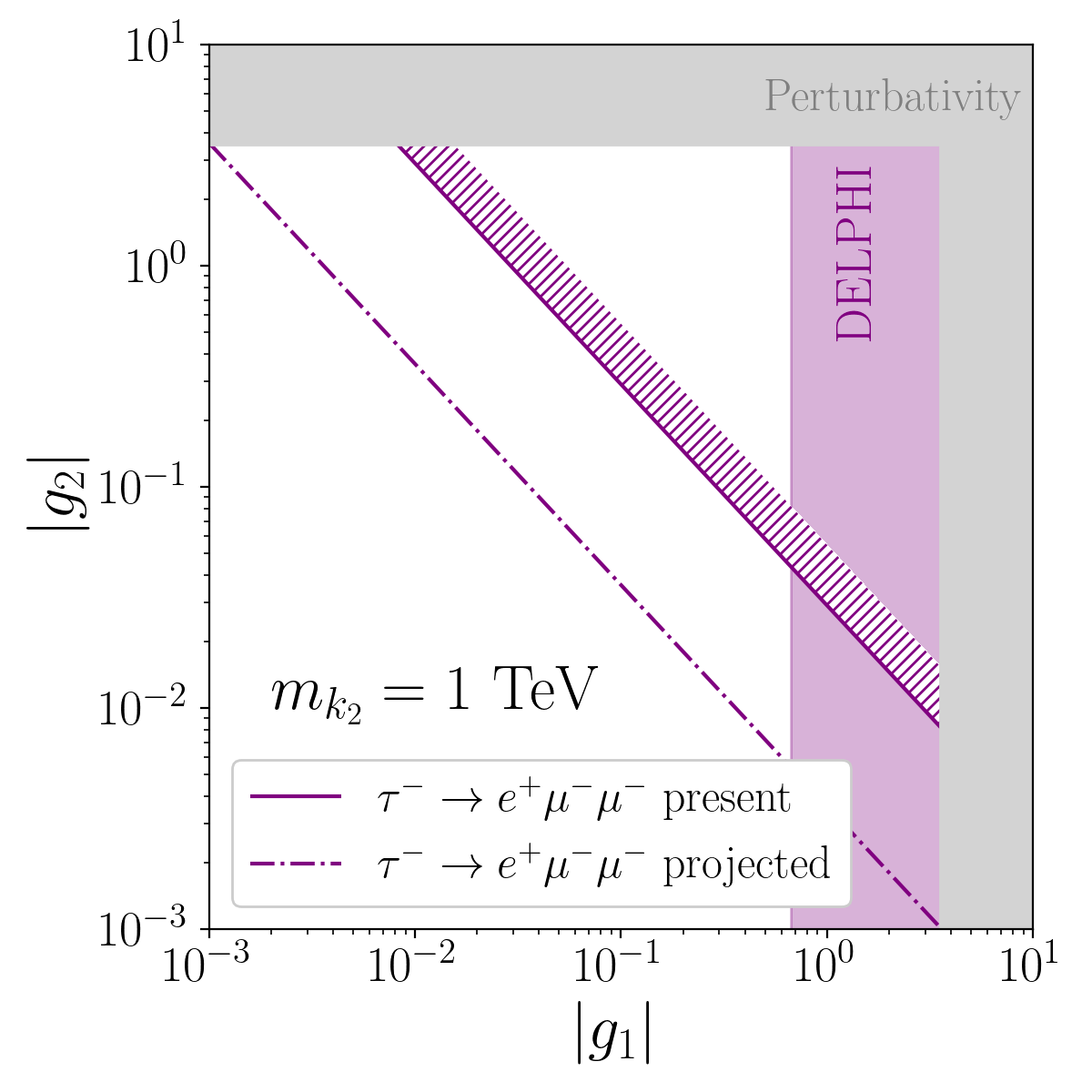}
    \includegraphics[width=0.48\textwidth]{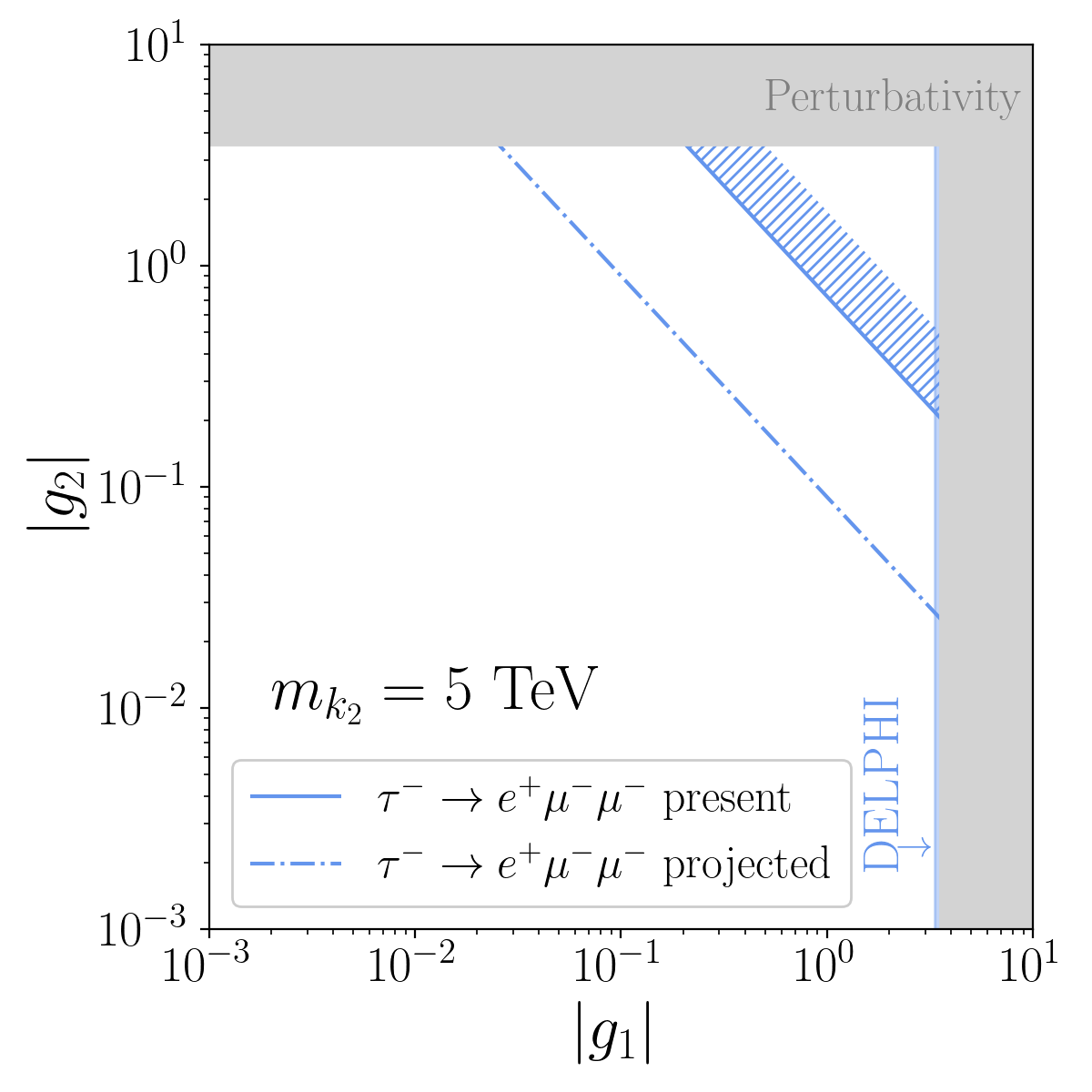}
    \caption{
    An overview of the electroweak singlet scalar parameter space. The present constraint on cLFV tau decays rules out the parameter space bounded by solid lines, in the direction of the hashing. The reach of Belle II is indicated by dot-dashed lines, i.e. observation of the decay $\tau^- \to \mu^+ e^- e^-$ ($\tau^- \to e^+ \mu^- \mu^-$) would be compatible with parameter space in the upper-right in the $|f_1|$-$|f_2|$ ($|g_1|$-$|g_2|$) plane above this line. The grey-shaded region is ruled-out by perturbativity of the BSM couplings.  The shaded region labelled `DELPHI' is ruled-out by the constraints in  Eq.~\eqref{eq:Delphi}, and where this is absent the constraint falls outside the perturbative regime.
    }
    \label{fig:singletbounds}
\end{figure}

Applying the present constraint quoted in Table~\ref{tab:LFVtau} and using Eq.~\eqref{eq:taubarmuee_k1}, the following upper-bound results:
\begin{align}
    \sqrt{|f_1 f_2|} \lesssim {0.17} 
    \ \frac{m_{k_1}}{\text{TeV}}.
    \label{eq:Singlet1-BRconstraint}
\end{align}
This constraint is shown  by the diagonal solid coloured lines in top panels of Fig.~\ref{fig:singletbounds} for the benchmark masses $m_{k_1} = 1$ and $5$ TeV. The coupling constant parameter space to the top-right of these lines is ruled out. The coloured dot-dashed lines show the expected reach of the Belle II experiment. The grey bands in Fig.~\ref{fig:singletbounds} display the regions for which the coupling constants are non-perturbative, and hence irrelevant for our analysis. 

The benchmark masses were chosen to be within the range that would produce an observable cLFV branching ratio at Belle II.  By saturating the perturbativity conditions (i.e.\ $f_1^2 = f_2^2 = 4\pi$), the projected sensitivity quoted in Table~\ref{tab:LFVtau} allows to probe $k_1$ masses:
\begin{align}
     m_{k_1} \lesssim {61}\;\;\text{TeV}.
     \label{eq:Singlet1-massreach}
\end{align}
Of course, for larger scalar masses the constraints considered become weaker, but so too does the observable effect in cLFV tau decays.  

\subsection{\texorpdfstring{$T=2$}{T=2} singlet model}
\label{subsec:T2Singlet}

For lepton triality $T=2$ the relevant Yukawa couplings in the Lagrangian of the doubly-charged scalar $k_2$ are obtained by exchanging the lepton flavours $1\leftrightarrow 2$ in the Lagrangian of Eq.~\eqref{eq:Lag0}, and are thus given by 
\begin{equation}\label{eq:Lag1}
    \mathcal{L}_{k_2} = \frac12 \left( 2 g_1 \overline{(\tau_{R})^c} e_{R} + g_2 \overline{(\mu_{R})^c} \mu_{R} \right) k_2 + \mathrm{h.c.}
\end{equation}
where we have set $g_{1,2}$ to be real and positive, as per the discussion in Sec.~\ref{subsec:T2Singlet}. 
The RH diagram in Fig.~\ref{fig:FeynmanDiagrams} leads to the decay modes $\tau^\pm \to \mu^\pm \mu^\pm e^\mp$. These may be parameterised by the LEFT Wilson coefficient
\begin{align}
    C^{\rm VRR}_{ee,2321} =\frac{g_1 g_2}{4m_{k_2}^2}
\end{align}
leading 
to the branching ratio 
\begin{align}
\label{eq:taubaremumu_k2}
   \mathrm{BR}(\tau^\pm \to  \mu^\pm \mu^\pm e^\mp) & =
\frac{g_1^2 g_2^2}{64G_F^2 m_{k_2}^4} \tilde{I}\left(\frac{m_\mu^2}{m_\tau^2}\right)\mathrm{BR}(\tau^- \to \mu^- \bar\nu_\mu \nu_\tau).
\end{align}
The effect of the muon mass appears in the factor $\tilde{I}(r)=  I_2(r)/I(r)$,
\begin{align}
    \label{eq:muonmassff}
I(r)=& 1-8r+8r^3-r^4-12r^2\ln(r) \\
I_2(r)=&\sqrt{1-4r}\left(1-2r(7+r+6r^2)\right)+24 r^2(1-r^2)\ln\left(\frac{1+\sqrt{1-4r}}{1-\sqrt{1-4r}}\right).
\end{align}
$I(r)$ is the usual muon mass effect in $\mathrm{BR}(\tau^- \to \mu^- \bar\nu_\mu \nu_\tau)$ and $I_2(r)$ the corresponding factor for $ \mathrm{BR}(\tau^\pm \to e^\mp \mu^\pm \mu^\pm)$, neglecting the electron mass. 

Similarly to the $k_1$ case, we may derive a bound on $k_2$ parameters using the present constraint quoted in Table~\ref{tab:LFVtau},
\begin{align}
    \sqrt{|g_1 g_2|} \lesssim 0.17\ \frac{m_{k_2}}{\text{TeV}}.
    \label{eq:Singlet2-BRconstraint}
\end{align}
This is plotted in the lower two panels of Fig.~\ref{fig:singletbounds} for the benchmark values of $m_{k_2}$. As with the $k_1$ case, the expected reach of Belle II is indicated by the dot-dashed lines. Fig.~\ref{fig:singletbounds} clearly demonstrates the strong constraints of cLFV leptonic $\tau$ decays on the electroweak singlet scalar and the improved sensitivity of the Belle II experiment.

The same benchmark masses are employed for the ${k_2}$ parameter study. Saturating the perturbativity conditions, the projected sensitivity in Table~\ref{tab:LFVtau} is able to probe models with an observable branching ratio at Belle II with 
\begin{align}
    m_{k_2} \lesssim {59}\ \text{TeV}.
    \label{eq:Singlet2-massreach}
\end{align}

\subsection{Direct searches}
\label{sec:DirectSearches}
The only available decay channels for the doubly-charged scalars are to pairs of same-sign leptons. ATLAS~\cite{ATLAS:2017xqs} searched for pair production of doubly-charged scalar singlets which subsequently decay to the $e^\pm e^\pm$, $e^\pm \mu^\pm$, or $\mu^\pm\mu^\pm$ same-sign dilepton final states. Their results provide the most stringent direct search constraints. Assuming similarly sized Yukawa couplings, the doubly-charged scalar $k_1$ ($k_2$) would have 50\% branching ratio to electrons (muons) and 50\% to $\tau\mu$ ($\tau e$) which maximises the product of the Yukawa couplings. So assuming that all events with a $\tau$ lepton in the final state are missed by the detector, there are lower bounds on the masses of the doubly-charged scalars given by
\begin{align}
m_{k_1} \geq 0.62\ \text{TeV},\qquad m_{k_2} \geq 0.57\ \text{TeV}, 
\end{align}
as per published data of Fig.~14 in Ref.~\cite{ATLAS:2017xqs}, rounded to two significant figures.
For a 100\% branching ratio to electrons (muons) the lower bounds become 
\begin{align}
m_{k_1} \geq 0.66\ \text{TeV},\qquad m_{k_2} \geq 0.72\ \text{TeV}. 
\end{align}

\subsection{Lepton scattering constraints}

The doubly-charged scalar also mediates $2\to 2$ scattering of leptons via t-channel exchange. In particular, $k_1$ contributes to $e^+ e^- \to e^+ e^-$ and $k_2$ contributes to $e^+ e^- \to \tau^+ \tau^-$, both of which have been constrained by the DELPHI experiment~\cite{DELPHI:2005wxt}. Translating the results in Ref.~\cite{Li:2019xvv}, we find the following lower limits on the $k_1$ and $k_2$ masses as a function of the Yukawa couplings with electrons,
\begin{align} 
\label{eq:Delphi}
    \frac{m_{k_1}}{|f_2|}  \geq 0.74\, \mathrm{TeV},
    \qquad
    \frac{m_{k_2}}{|g_1|}  \geq 1.5\, \mathrm{TeV} .
\end{align}
These constraints are indicated by coloured bands in Fig.~\ref{fig:singletbounds}. For larger $m_{k_{1,2}}$ masses, the DELPHI constraint lies outside of the perturbative regime for the coupling constants.

\subsection{Other observables}

Other observables include leptonic Higgs and $Z$ boson decays.  We find that the most striking signal would be the cLFV $Z$ decay to four leptons.\footnote{Decays of the Higgs and $Z$ bosons to two charged leptons receive corrections at 1-loop level, which is left for future work.} The dominant contribution to these comes from decays to $\tau^+ \tau^-$ followed by a cLFV $\tau$ decay to three leptons. Higgs decays are not as sensitive as they are suppressed by the $\tau$ Yukawa coupling, and thus we focus on $Z$ boson decays.

\subsubsection{Flavour-violating \texorpdfstring{$Z$}{Z} decays}

There are two contributions to cLFV $Z$ boson decays: the decay via two off-shell scalar electroweak singlets, or $Z$ decays to $\tau^+\tau^-$ followed by a leptonic cLFV $\tau$ decay. The former is highly suppressed due to the constraint on the electroweak singlet scalar mass. For example, for the scalar ${k_1}$, this can be seen from the branching ratio
\begin{align}
    \mathrm{BR}(Z\to k_1^{++} k_1^{--} \to  e^+ e^+  \mu^- \tau^-) & = \frac{f_1^2 f_2^2 \alpha_{\rm em} }{2^{15}\cdot 3 \cdot 5^2\cdot 7 \pi^4} \frac{(T_3 -Q s_w^2)^2}{s_w^2 c_w^2} \frac{m_Z^9}{m_{k_1}^8 \Gamma_Z} ,
\end{align}
where $Q$ denotes the electric charge of the scalar, and $T_3$ its third component of weak-isospin. Therefore it is justified to neglect this contribution, and to approximate cLFV $Z$ boson decays by
\begin{equation}
\begin{aligned}\label{eq:Z-4lepton}
    \text{BR}(Z \to \tau^+\tau^-\to e^+ e^+ \mu^-\tau^-)&=  \text{BR}(Z \to \tau^+\tau^-)\cdot \text{BR}(\tau^+\to e^+ e^+\mu^-), \\
    &=0.036\, \text{BR}(\tau^+\to e^+ e^+ \mu^-).
\end{aligned}
\end{equation}
A similar argument can be made for $k_2$. Searching for the cLFV $Z$ boson decay $Z\to e^+ e^+ \mu^- \tau^-$ thus provides an interesting probe, directly related to the cLFV in $\tau$ decays at the focus of this work. Conversely, constraining the cLFV $\tau$ decays also provides indirect constraints for cLFV $Z$ boson decays. As the upper limits for these $\tau$ decays are generally stronger than those for the corresponding $Z$ boson decays, we do not expect competitive constraints from the latter.

\subsubsection{Anomalous magnetic moments}

For completeness, we note that the doubly-charged scalar also contributes to lepton anomalous magnetic moments~\cite{Li:2018cod,Li:2019xvv} through
\begin{equation}
    0\leq \Delta a_\ell  = \frac{m_\ell^2}{24 \pi^2 m_{k_T}^2} \begin{cases}
        f_1^2 & \text{for}\ T=1, \\
        g_2^2 & \text{for}\ T=2. \\
    \end{cases} 
\end{equation}
This effect is too small to be of practical phenomenological interest.

\section{Electroweak triplet models}
\label{sec:triplet}

The construction above can be mirrored to produce alternative models where the doubly-charged scalar is embedded in a $Y=1$ weak-isospin triplet, $\Delta_T$, where (as before) $T$ denotes lepton triality. As for the singlet case, only $T=1,2$ are relevant for models of cLFV $\tau$ decays. The triplet models have richer phenomenology than the singlet models, due mainly to including effects of the weak-isospin partners of the doubly-charged scalars.

As well as the doubly-charged scalar $\Delta_T^{++}$, such a triplet also contains a singly-charged scalar $\Delta_T^+$ and a neutral complex scalar $\Delta_T^0$. It is convenient to represent this complex triplet using a traceless $2 \times 2$ matrix of the form
\begin{align}
    \mathbf{\Delta}_T = \left( \begin{array}{cc} \frac{\Delta_T^+}{\sqrt{2}} &\ \ \Delta_T^{++} \\
    \Delta_T^0 & \ \ -\frac{\Delta_T^+}{\sqrt{2}} \end{array} \right) = \Delta_{T}^i \frac{\sigma_i}{\sqrt{2}},
\end{align}
where $\sigma_i$ denotes the Pauli matrices.
A weak-isospin transformation is represented through
\begin{align}
    \mathbf{\Delta}_T \to U \mathbf{\Delta}_T U^\dagger
\end{align}
where $U$ is in the fundamental representation of SU(2). The normalisations have been chosen so that the quadratic invariant
\begin{align}
    \text{tr}(\mathbf{\Delta}_T^\dagger \mathbf{\Delta}_T) = \Delta_T^{--} \Delta_T^{++} + \Delta_T^{-} \Delta_T^{+} + \Delta_T^{0*} \Delta_T^{0}
\end{align}
produces standard normalisations for the component fields.

\subsection{\texorpdfstring{$T=1$}{T=1} triplet model}

For lepton triality $T=1$ the relevant Yukawa coupling Lagrangian is
 \begin{equation}
    \mathcal{L}_{\Delta_1} = \frac12 \left( 2 f_1\, \overline{L_3^c}\ i\sigma_2 \mathbf{\Delta}_1\, L_2 + f_2 \, \overline{L_1^c}\ i\sigma_2 \mathbf{\Delta}_1\, L_1 \right) + \mathrm{h.c.}
\end{equation}
where
\begin{align}
     \overline{L_3^c}\ i\sigma_2 \mathbf{\Delta}_1\, L_2 &= - \overline{(\tau_L)^c}\, \mu_L\, \Delta_1^{++} - \frac{1}{\sqrt{2}} \left[ \overline{(\tau_L)^c}\, \nu_{\mu L} +  \overline{(\nu_{\tau L})^c}\, \mu_L \right] \Delta_1^+  + \overline{(\nu_{\tau L})^c}\, \nu_{\mu L}\, \Delta_1^0, \label{eq:Delta1-Yuk1}\\
     \overline{L_1^c}\ i\sigma_2 \mathbf{\Delta}_1\, L_1 &= - \overline{(e_L)^c}\, e_L\, \Delta_1^{++} - \sqrt{2}\, \overline{(e_L)^c}\, \nu_{e L}\, \Delta_1^+  + \overline{(\nu_{e L})^c}\, \nu_{e L}\, \Delta_1^0.
     \label{eq:Delta1-Yuk2}
\end{align}
Similarly to the singlet models, the phases of these Yukawa coupling constants can be absorbed by field redefinitions, so we set them to be real-valued and positive from now on without loss of generality. 
Note that this is the same triplet that appears in the type-II seesaw mechanism~\cite{Konetschny:1977bn,Magg:1980ut,Cheng:1980qt,Schechter:1980gr}, whose contribution to neutrino masses is briefly discussed in Sec.~\ref{sec:neutrino}.

\begin{figure}[t!]
    \centering
    \includegraphics[width=0.48\textwidth]{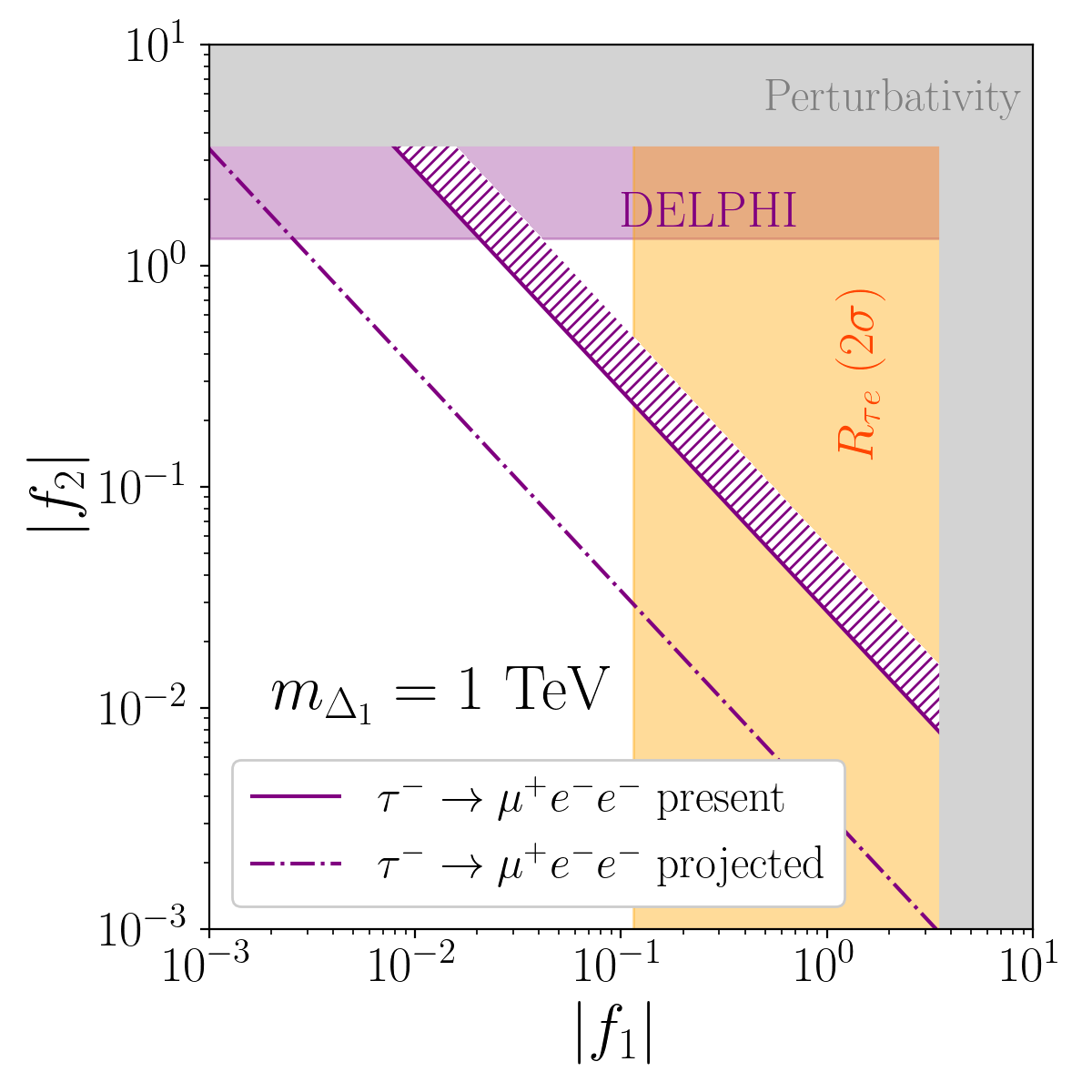}
     \includegraphics[width=0.48\textwidth]{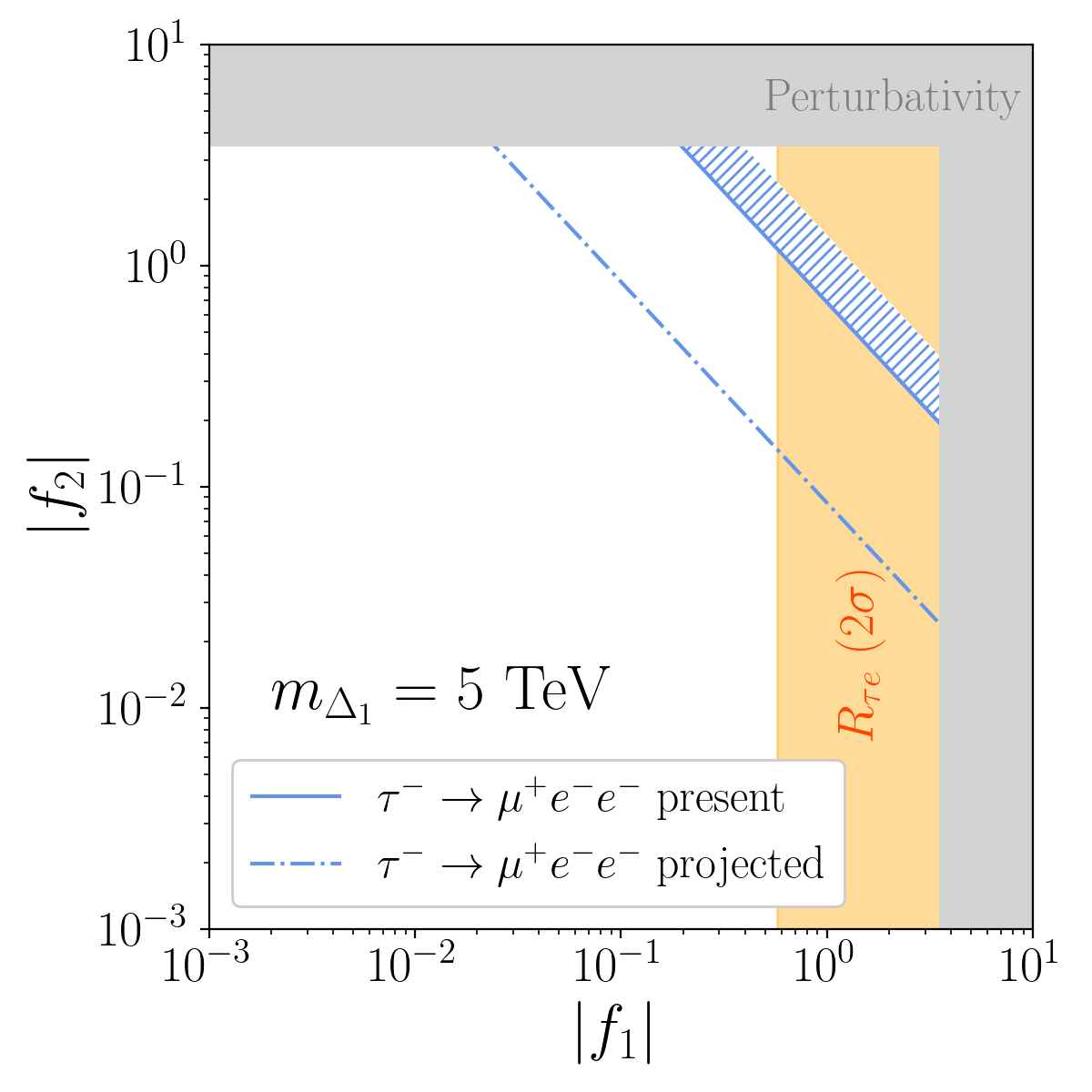}
       \includegraphics[width=0.48\textwidth]{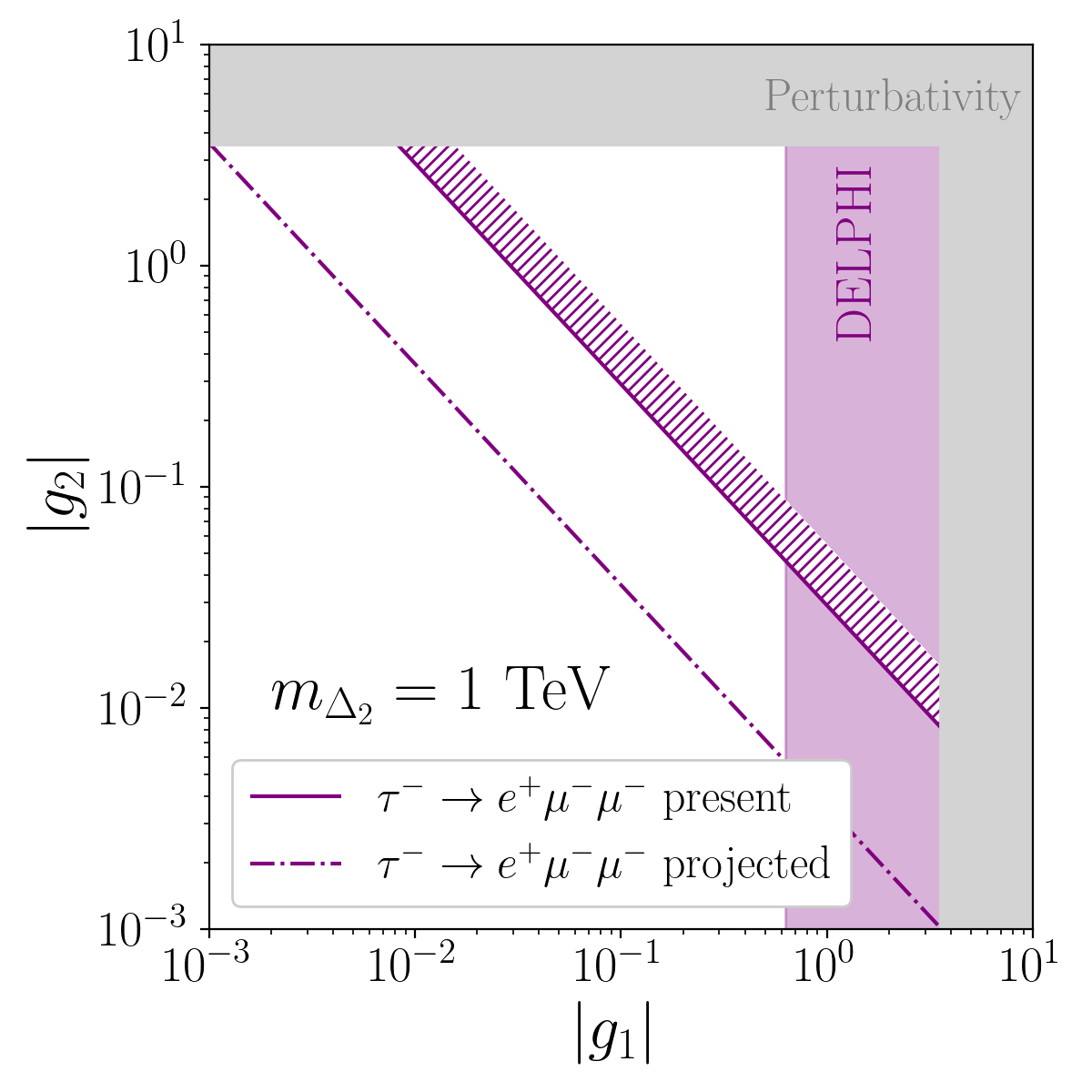}
     \includegraphics[width=0.48\textwidth]{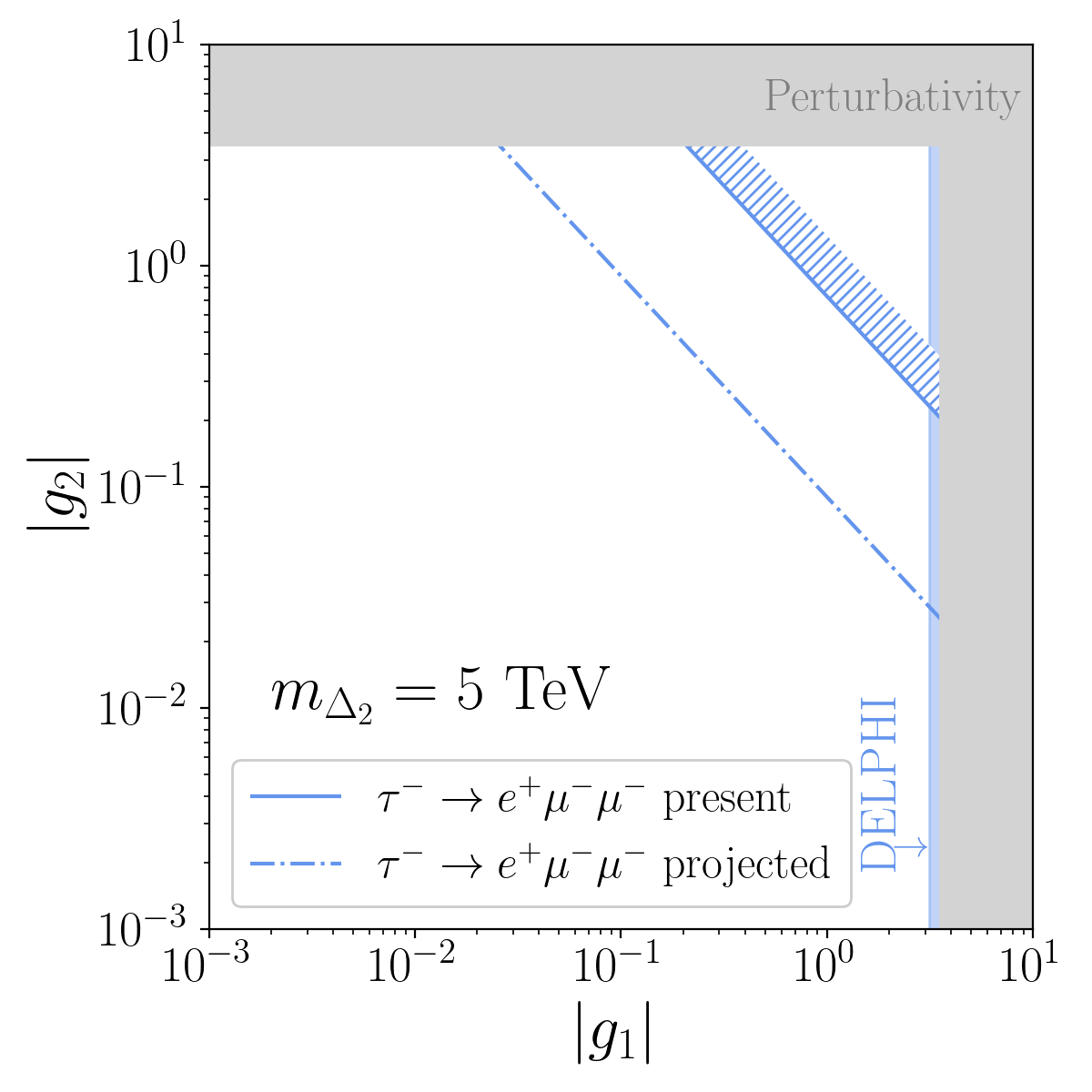}
    \caption{ 
       An overview of the electroweak triplet scalar parameter space. The present constraint on cLFV tau decays rules out the parameter space bounded by solid lines, in the direction of the hashing. The reach of Belle II is indicated by dot-dashed lines, i.e. observation of the decay $\tau^- \to \mu^+ e^- e^-$ ($\tau^- \to e^+ \mu^- \mu^-$) would be compatible with parameter space in the upper-right in the $|f_1|$-$|f_2|$ ($|g_1|$-$|g_2|$) plane above this line. The grey-shaded region is ruled-out by perturbativity of the BSM couplings.  The shaded region labelled `DELPHI' is ruled-out by the constraints in  Eq.~\eqref{eq:DELPHItriplet}, and where this is absent the constraint falls outside the perturbative regime. For the lepton universality ratios double ratios $R_{\ell_i \ell_j}$ of leptonic $\tau$ decays, the shaded region is the constraint derived when theory is required to match experiment at the $2\sigma$ level, as labelled. For $\Delta_2$ in the lower row, these constraints are not shown because they are weaker than those from DELPHI.
}
    \label{fig:tripletbounds}
\end{figure}

For energies below the mass of $\Delta_1$, the Wilson coefficient\footnote{
The matching of the electroweak triplet scalar model has been derived in Refs.~\cite{Du:2022vso,Li:2022ipc} up to 1-loop order.
} relevant for the $\tau^\pm \to  \mu^\mp e^\pm e^\pm$ decays is
\begin{align}
C^{\rm VLL}_{ee,1213} & = \frac{f_1 f_2}{4 m_{\Delta_1}^2}
\end{align}
with branching ratios given by
\begin{align}
    \mathrm{BR}(\tau^\pm \to \mu^\mp e^\pm e^\pm ) & =\frac{f_1^2f_2^2}{64 G_F^2 m_{\Delta_1}^4 }  \mathrm{BR}(\tau^- \to \mu^- \bar\nu_\mu \nu_\tau).
\end{align}
The current constraint from the non-observation of these decays is identical to the $k_1$ singlet cases, i.e. Eq.~(\ref{eq:Singlet1-BRconstraint}) holds with $m_{k_1}$ replaced by $m_{\Delta_1}$
\begin{align}
    \sqrt{|f_1 f_2|} \lesssim {0.17} 
    \ \frac{m_{\Delta_1}}{\text{TeV}}.
    \label{eq:Triplet1-BRconstraint}
\end{align}
Similarly, the projected Belle II mass reach is roughly {$61$} TeV. These constraints are depicted by the diagonal solid coloured lines in the top row of Fig.~\ref{fig:tripletbounds} for three benchmark masses, with the region to the top-right being ruled out. The projected reach of Belle II is indicated by the dot-dashed lines. The same benchmark masses are studied here as were studied for the electroweak singlet scalar models.

\subsection{\texorpdfstring{$T=2$}{T=2} triplet model}

Taking the electroweak triplet scalar with lepton triality $T=2$, the Yukawa coupling Lagrangian for $\Delta_2$ is  
\begin{equation}
    \mathcal{L}_{\Delta_2} = \frac12 \left( 2 g_1 \overline{L_3^c}\, i\sigma_2 \mathbf{\Delta}_2\, L_1 + g_2 \overline{L_2^c}\, i\sigma_2 \mathbf{\Delta}_2\, L_2 \right) + \mathrm{h.c.}
\end{equation}
with real and positive Yukawa couplings $g_{1,2}$.
These expand out to give Eq.~(\ref{eq:Delta1-Yuk1}) with the substitution $\mu \to e$ for the $g_1$ term, and Eq.~(\ref{eq:Delta1-Yuk2}) with $e \to \mu$ for the $g_2$ term:
\begin{align}
     \overline{L_3^c}\ i\sigma_2 \mathbf{\Delta}_2\, L_1 &= - \overline{(\tau_L)^c}\, e_L\, \Delta_2^{++} - \frac{1}{\sqrt{2}} \left[ \overline{(\tau_L)^c}\, \nu_{e L} +  \overline{(\nu_{\tau L})^c}\, e_L \right] \Delta_2^+  + \overline{(\nu_{\tau L})^c}\, \nu_{e L}\, \Delta_2^0, \label{eq:Delta2-Yuk1}\\
     \overline{L_2^c}\ i\sigma_2 \mathbf{\Delta}_2\, L_2 &= - \overline{(\mu_L)^c}\, \mu_L\, \Delta_2^{++} - \sqrt{2}\, \overline{(\mu_L)^c}\, \nu_{\mu L}\, \Delta_2^+  + \overline{(\nu_{\mu L})^c}\, \nu_{\mu L}\, \Delta_2^0.
     \label{eq:Delta2-Yuk2}
\end{align}

The Wilson coefficient
\begin{align} 
C^{VLL}_{ee,2321} & = \frac{g_1 g_2}{4m_{\Delta_2}^2}
\end{align}
parametrises the strength of the decays $\tau^\pm \to  \mu^\pm \mu^\pm e^\mp$, with the branching ratio being
\begin{align}
    \mathrm{BR}(\tau^\pm \to \mu^\pm \mu^\pm e^\mp) & =\frac{g_1^2g_2^2}{64 G_F^2 m_{\Delta_2}^4 } \tilde I\left(\frac{m_\mu^2}{m_\tau^2}\right) \mathrm{BR}(\tau^- \to \mu^- \bar\nu_\mu \nu_\tau).
\end{align}

As for the $\Delta_1$ model, the current constraint from the non-observation of these decays is the same as for the singlet $k_2$ case, i.e.  Eq.~(\ref{eq:Singlet2-BRconstraint}) holds with the $m_{k_2}$ replaced by $m_{\Delta_2}$, 
\begin{align}
    \sqrt{|g_1 g_2|} \lesssim 0.17\ \frac{m_{\Delta_2}}{\text{TeV}}
    \label{eq:Triplet2-BRconstraint}
\end{align}
and the Belle II mass reach is roughly {$59$} TeV. This bound and the Belle II reach are illustrated in the bottom row of Fig.~\ref{fig:tripletbounds}.

\subsection{Direct searches}
\label{sec:triplet_direct}
As for the singlet cases, there are constraints from direct searches for doubly-charged scalars decaying to a pair of same-sign leptons. A difference between the singlet and triplet models is that the former involve RH leptons, while the latter feature LH leptons. From the published data of Fig.~13 in Ref.~\cite{ATLAS:2017xqs} we infer lower limits on the masses of the doubly-charged scalars $\Delta_{1,2}^{++}$ given by
\begin{align}
    m_{\Delta_1} \geq 0.69\ \text{TeV}\quad \text{and}\quad m_{\Delta_2} \geq 0.73\ \text{TeV}
\end{align}
for the case of 50\% BR to electrons and muons, respectively, with the final states containing a $\tau$ which is assumed to be unobserved. For a 100\% branching ratio to electrons (muons) the lower bounds become 
\begin{align}
m_{\Delta_1} \geq 0.77\ \text{TeV}\quad \text{and}\quad m_{\Delta_2} \geq 0.85\ \text{TeV} 
\end{align}
according to the published data of Fig.~10 in Ref.~\cite{ATLAS:2017xqs}.

\subsection{Lepton scattering constraints}

As for the singlet cases, the DELPHI experiment~\cite{DELPHI:2005wxt} places constraints on the doubly-charged scalar mass $m_{\Delta_1}$ and coupling constant $f_2$ from $e^+ e^- \to e^+ e^-$, and $m_{\Delta_2}$ and $g_1$ from $e^+ e^- \to \tau^+ \tau^-$. Translating the results in Ref.~\cite{Li:2019xvv}, we obtain the bounds
\begin{align} 
\label{eq:DELPHItriplet}
    \frac{m_{\Delta_1}}{|f_2|} & \geq 0.75\, \mathrm{TeV},
    &
    \frac{m_{\Delta_2}}{|g_1|} & \geq 1.6\, \mathrm{TeV} .
\end{align}
Note that these constraints are very slightly different from the analogous singlet bounds because the opposite chiral structure changes the details of the interference between the SM and triplet-exchange diagrams.

\subsection{Leptonic processes involving neutrinos}

The electroweak triplet scalars introduce new contributions to leptonic processes with neutrinos. We follow the discussion in Refs.~\cite{Li:2018cod,Li:2019xvv,Li:2021lnz} and focus on the most stringent electroweak physics constraints~\cite{Li:2018cod,Li:2019xvv}, namely lepton flavour universality, neutrino trident and shifts of the Fermi constant and its impact on the weak mixing angle, the $W$ boson mass and CKM unitarity. The latter three do not receive direct contributions at tree level, but are modified indirectly via the shift of the Fermi constant extracted in muon decay, $G_{F, \mu}$.

In particular, the partial width for the decay $\ell_\alpha\to \ell_\beta \nu_i\bar \nu_j(\gamma)$ in terms of relevant Wilson coefficients is given by~\cite{Kinoshita:1958ru,Marciano:1988vm,Fael:2013pja,Ferroglia:2013dga}
\begin{equation}
    \Gamma(\ell_\alpha\to \ell_\beta \nu_i \bar\nu_j(\gamma)) = \frac{G_F^2 m_\alpha^5}{192\pi^3} I\left(\frac{m_\beta^2}{m_\alpha^2}\right) R_W R_\gamma \sum_{i,j}\frac{|C^{VLL}_{\nu e,ij \beta\alpha}|^2+ |C^{VLR}_{\nu e,ij \beta\alpha}|^2}{8 G_F^2},
\end{equation}
where the functions $R_W$ and $I$ parameterise corrections from finite lepton masses, while $R_\gamma$ parametrises the emission of soft photons. The function $I$ is defined in Eq.~\eqref{eq:muonmassff} and the other functions are given by
\begin{align}
    R_W  = 1+ \frac35 \frac{m_\alpha^2}{m_W^2}+ \frac95 \frac{m_\beta^2}{m_W^2},\qquad 
    R_\gamma  = 1+ \frac{\alpha(m_\alpha)}{2\pi}\left(\frac{25}{4}-\pi^2\right)
    \nonumber \\
    \quad \text{with}\quad
    \alpha(m_\alpha)^{-1} = \alpha^{-1} - \frac{2}{3\pi} \ln\left(\frac{m_\alpha}{m_e}\right) + \frac{1}{6\pi}.
\end{align}
The SM prediction for the LEFT Wilson coefficients is
\begin{align}\label{eq:nu-e-SM}
    C^{\rm VLL,SM}_{\nu e, ij \beta\alpha} & =-2\sqrt{2} G_F \left(-\frac12 + s^2_w\right)\delta_{ij}\delta_{\alpha\beta} -2\sqrt{2} G_F\delta_{i\alpha}\delta_{j\beta} ,
    & 
    C^{\rm VLR,SM}_{\nu e, ij \beta\alpha} & = -2\sqrt{2} G_F s_w^2 \delta_{ij} \delta_{\alpha\beta}
    .
\end{align}
The electroweak triplet contributes via the 4-lepton SMEFT operators $O^{ll}$ to $C^{VLL}_{\nu e}$ as described by the LEFT matching conditions in App.~\ref{sec:LEFT} and thus modify the leptonic muon and tau decays.

\subsubsection{Lepton flavour universality}

The most direct probe is provided by the lepton flavour universality double ratios,
\begin{align}
    R_{\mu e} & = \frac{\Gamma(\tau\to\mu +\mathrm{inv})}{\Gamma(\tau\to e + \mathrm{inv})}
    \frac{\Gamma_{\rm SM}(\tau\to e +\mathrm{inv})}{\Gamma_{\rm SM}(\tau\to \mu + \mathrm{inv})},
    \\
    R_{\tau \mu} & = \frac{\Gamma(\tau\to e +\mathrm{inv})}{\Gamma(\mu\to e + \mathrm{inv})}
    \frac{\Gamma_{\rm SM}(\mu\to e +\mathrm{inv})}{\Gamma_{\rm SM}(\tau\to e + \mathrm{inv})},
    \\
    R_{\tau e} & = \frac{\Gamma(\tau\to \mu +\mathrm{inv})}{\Gamma(\mu\to e + \mathrm{inv})}
    \frac{\Gamma_{\rm SM}(\mu\to e +\mathrm{inv})}{\Gamma_{\rm SM}(\tau\to \mu + \mathrm{inv})}
    ,
\end{align}
of which only two are independent. For the triplet $\Delta_1$ we find
    \begin{align}
     R_{\tau e} = R_{\mu e} = \frac{\left(1 - \frac{f_1^2}{8\sqrt{2} G_F m_{\Delta_1}^2}  \right)^2}{1+ \frac{(f_1f_2)^2}{2^{7} G_F^2 m_{\Delta_1}^4}}
     \approx 1-\frac{f_1^2}{4\sqrt{2}G_F m_{\Delta_1}^2},\qquad
    R_{\tau\mu} = 1
    \end{align}
    and for $\Delta_2$
    \begin{align}
    R_{\tau\mu}  = R_{\mu e}^{-1} = \frac{\left(1-\frac{g_1^2}{8\sqrt{2} G_F m_{\Delta_2}^2} \right)^2} {1+\frac{(g_1 g_2)^2}{2^{7} G_F^2 m_{\Delta_2}^4}}
     \approx 1-\frac{g_1^2}{4\sqrt{2}G_F m_{\Delta_2}^2},\qquad
     R_{\tau e}  = 1
     ,
  \end{align}
  where we neglected terms quartic in the Yukawa couplings in the approximation. 
  
Using the experimental values in Ref.~\cite{ParticleDataGroup:2020ssz}, we find for the double ratios~\cite{Li:2019xvv}
\begin{align}
R_{\mu e} = 1.0034(32), \hspace{0.5cm} R_{\tau \mu} = 1.0022(29),\hspace{0.5cm} R_{\tau e} = 1.0056(29). \label{eq:LEFUratios}
\end{align}
Each of these is to be compared to the SM prediction of unity. We note that there is a tension with the SM prediction for $R_{\tau e}$  which cannot be alleviated at the central value by the electroweak triplet scalar. 
Requiring that the model agrees with the values in Eq.~\eqref{eq:LEFUratios} to within $2\sigma$, the strongest constraints from these ratios come from $R_{\tau e}$ for $\Delta_1$ and $R_{\tau \mu}$ for $\Delta_2$. The excluded regions are indicated by the vertical coloured bands in the top row of Fig.~\ref{fig:tripletbounds}. The region excluded by $R_{\tau\mu}$ is not shown in the bottom row because $|g_1|$ is more strongly constrained by the DELPHI measurement.

\subsubsection{Trident process}

The singly-charged scalar $\Delta_2^+$ contributes at tree-level via the $g_2$ Yukawa interaction to the trident process $\nu_\mu N \to \nu_\mu \mu^+\mu^- N$, where $N$ represents a nucleon. This is parameterised by the Wilson coefficient
\begin{align}
    C^{\rm VLL,NP}_{\nu e,2222} 
    = \frac{g_2^2}{4 m_{\Delta_2}^2},
\end{align}
which gives the ratio of the modified cross section $\sigma$ to the SM cross section $\sigma_{\text{SM}}$ to be
\begin{equation}
\frac{\sigma}{\sigma_{\text{SM}}} = \frac{ (1+4\sin^2\theta_w -  C^{\rm VLL,NP}_{\nu e,2222}/(\sqrt{2}G_F))^2 + ( 1 - C^{\rm VLL,NP}_{\nu e,2222}/ (\sqrt{2}G_F))^2}{(1+4 \sin^2\theta_w)^2+ 1}.
\end{equation}
As the Wilson coefficient $C^{\rm VLL,NP}_{e\nu,2222}$ is strictly positive, the electroweak triplet contribution reduces the ratio to be less than one.

The CHARM-II~\cite{CHARM-II:1990dvf}, CCFR~\cite{CCFR:1991lpl} and NuTeV~\cite{NuTeV:1999wlw} experiments have measured this ratio obtaining $1.58 \pm 0.64$, $0.82 \pm 0.28$ and $0.72^{+1.73}_{-0.72}$, respectively. In order to estimate how neutrino trident production $\nu_\mu N \to \nu_\mu \mu^+ \mu^- N$ constrains the electroweak triplet scalar $\Delta_2$, we combine the experimental measurements\footnote{For the combination we assume Gaussian distributions and utilise the larger upper error for the NuTeV result.} to obtain $0.94 \pm 0.25$, which constrains
\begin{align}
-0.15\leq\frac{C^{\rm VLL,NP}_{\nu e,2222}}{\sqrt{2} G_F} \leq 0.28
\qquad\implies\qquad
|g_2| \lesssim 4.3\, \frac{m_{\Delta_2}}{\text{TeV}}.
\end{align}
This may be compared with the bound in Eq.~(\ref{eq:Triplet2-BRconstraint}).
Numerically it is a weak constraint, with the upper bound on $g_2$ in the non-perturbative regime and thus moot. However, it is worth noting that the constraint is purely on $g_2$ rather than the product $g_1 g_2$, and so in principle the trident process provides a complementary constraint.

As this constraint is rather weak in this model (and also orthogonal to the main discussion of cLFV leptonic $\tau$ decays) we do not discuss the sensitivity of neutrino trident processes at DUNE. DUNE is able to measure other trident processes, including lepton-flavour-violating neutrino trident processes. See Refs.~\cite{Ballett:2018uuc,Altmannshofer:2019zhy} for a detailed discussion of neutrino tridents at DUNE.

\subsubsection{Fermi constant}
\begin{table}[t!]
    \centering
\setlength{\tabcolsep}{10pt}
\renewcommand{\arraystretch}{1.5}
	\begin{tabular}{|c|c|}
 \hline
Input & Value\\
\hline 
         $m_Z$ & $91.1876(21)$ GeV\\
         $G_{F,\mu}$  & $1.1663787(6) \times 10^{-5}$ GeV$^{-2}$\\
         $\alpha^{-1}$ & 137.035999180(10)\\
         \hline 
         \end{tabular}
    \caption{Input parameters for electroweak observables, taken from Ref.~\cite{ParticleDataGroup:2020ssz}.}
    \label{tab:input}
\end{table}

These models generate new contributions to muon decay, which is used to determine the Fermi constant $G_F$. We denote the Fermi constant determined through muon decay $G_{F,\mu}$.
Using the input parameters for the SM electroweak observables listed in Table~\ref{tab:input}, these contributions introduce a shift in $G_{F,\mu}$, 
\begin{equation}
G_{F,\mu}^2=\frac{1}{8}\sum_{i,j}\left(|C^{VLL}_{\nu e,ij12}|^2+ |C^{VLR}_{\nu e,ij 12}|^2\right) ,
\end{equation}
in contrast to the SM value $G_{F,0}=(\sqrt{2} v^2)^{-1/2}$. Here $v$ is the vacuum expectation value~(VEV) of the Higgs field $\left\langle H \right\rangle=v/\sqrt{2}$. We define $G_{F,0} = G_{F,\mu}(1+\delta G_F)$, and for $\Delta_1$ and $\Delta_2$ we find that
\begin{align}
\delta G_F &= \begin{cases}
-  \frac{f_1^2 f_2^2}{1024 G_{F,\mu}^2 m_{\Delta_1}^4}  & \mathrm{for}\;\Delta_1,\\[2ex]
-  \frac{g_1^2 g_2^2}{1024 G_{F,\mu}^2 m_{\Delta_2}^4}  & \mathrm{for}\;\Delta_2.
\end{cases}
\end{align}
As the correction to the Fermi constant only occurs at quartic order in the Yukawa couplings, we do not expect strong constraints from $\delta G_F$. 

Several measurements are sensitive to the Fermi constant and thus provide constraints on $\delta G_F$. Here we consider the weak mixing angle, the $W$ boson mass and CKM unitarity as a subset of these measurements. For each of these, we use the tree-level SM expression to determine how the shift in $G_F$ affects the observable, and then derive a constraint on the shift $\delta G_F$ the electroweak fit from Ref.~\cite{deBlas:2021wap} (which includes loop-level SM corrections). This fit provides SM predictions for the different observables (without including the measurements) which are then compared to the experimental measurements to obtain constraints on $\delta G_F$.

The effective leptonic weak mixing angle does not receive direct corrections at tree-level\footnote{We neglect loop-level corrections to $\Pi_{\gamma Z}(m_Z^2)$ in the analysis.} because the $Z$ boson couplings to leptons are not modified. Therefore, we find
\begin{align}
    \bar s_{\ell}^2 & = s_{w,0}^2 \left(1- \frac{1-s_{w,0}^2}{1-2s_{w,0}^2} \delta G_F\right),
\end{align}
which has been obtained from the tree-level SM prediction of the weak mixing angle 
\begin{equation}
    s_{w,0}^2 = \frac12 \left(1-\sqrt{1-\frac{4\pi \alpha}{\sqrt{2} G_{F} m_Z^2 
    }}\right) .
 \end{equation}
Contrasting this with the experimental result $\bar s_{\ell,\rm exp}^2 (\mathrm{LEP}) = 0.23153(4)$~\cite{ALEPH:2005ab} with the loop-corrected SM prediction for the weak mixing angle based on the fit in \cite{deBlas:2021wap}, $\bar s_{\ell,\rm SM}^2  = 0.231534(41)$, results in 
\begin{align}
\label{eq:DeltaGF_constraint}
   \delta G_F = 12 (170)\times 10^{-6} . 
\end{align}

A shift in the Fermi constant also translates to a shift in the $W$ boson mass
\begin{equation}
    m_W^2 = m_{W,0}^2\left(1+\frac{s_{w,0}^2}{1-2s_{w,0}^2}\delta G_F\right)
\end{equation}
which has been derived from the tree-level SM prediction
\begin{equation}
    m_{W,0} = m_Z \sqrt{1-s_{w,0}^2} .
\end{equation}
A comparison of the experimental global fit to the $W$ boson mass excluding the new CDF measurement 
$m_{W,\rm exp}=80.377(12)$~\cite{ParticleDataGroup:2020ssz} [and a combination of all Tevatron measurements by themselves $m_{W,\rm Tevatron}=$80.4274(89)~\cite{CDF:2022hxs}]
with the SM prediction $m_W=80.3545(42)$ GeV~\cite{deBlas:2021wap} results in 
\begin{align}
   \delta G_F = 0.00130(73) \, [ 0.00421(57)  ].
\end{align}

A shift in the Fermi constant also leads to an apparent violation of CKM unitarity. We find for the unitarity relation of the first row CKM matrix elements that
\begin{equation}
    \sum_\beta |V_{u\beta}|^2 = 1+2 \delta G_F.
\end{equation}
The global fit in Ref.~\cite{ParticleDataGroup:2020ssz} requires $\sum_\beta |V_{u\beta}|^2 = 0.9985\pm0.0007$. If we conservatively demand consistency at $3\sigma$ (to include the SM prediction), we find
\begin{align}
   \delta G_F = -0.00075(35)  .
\end{align}

Note that the different observables are in tension with each other: the leptonic weak mixing angle prefers no correction to $\delta G_F$, the $W$ boson mass indicates a positive $\delta G_F$, and CKM unitarity negative $\delta G_F$.
As $\delta G_F$ in the electroweak triplet model is proportional to the fourth power of Yukawa couplings, none of the observables constraining $\delta G_F$ provide a competitive constraint. Even the sensitivity of the leptonic weak mixing angle (which probes $\delta G_F$ at the level of $10^{-4}$) is only sensitive to scales $m_{\Delta_T}\lesssim 0.50$ TeV for Yukawa couplings of order unity, which is already excluded by direct searches at the LHC~(see Sec.~\ref{sec:triplet_direct}).

\subsection{Other observables}
As discussed for the electroweak singlet scalar, there are new contributions to leptonic Higgs and $Z$ boson decays and the anomalous magnetic moment~\cite{Li:2018cod,Li:2019xvv}. These are too small to have any measurable phenomenological implications.

\section{Phase space}
\label{sec:phase space}

In the caption of Table~\ref{tab:LFVtau}, we alluded to upper limits placed by experiment on cLFV leptonic $\tau$ decays depending on the assumed distribution of signal events. The limits quoted in this table are extracted by the experimental collaborations assuming that the leptons from the $\tau$-decay follow a phase space distribution, i.e.~no kinematic dependence in the matrix element.
The model-discriminating power of three-body phase space in tau decays has been studied, for example, in Refs.~\cite{Dassinger:2007ru,Goto:2010sn,Celis:2014asa,Bruser:2015yka}. 

\begin{figure}[t!]
    \centering  
    \includegraphics[width=0.95\textwidth]{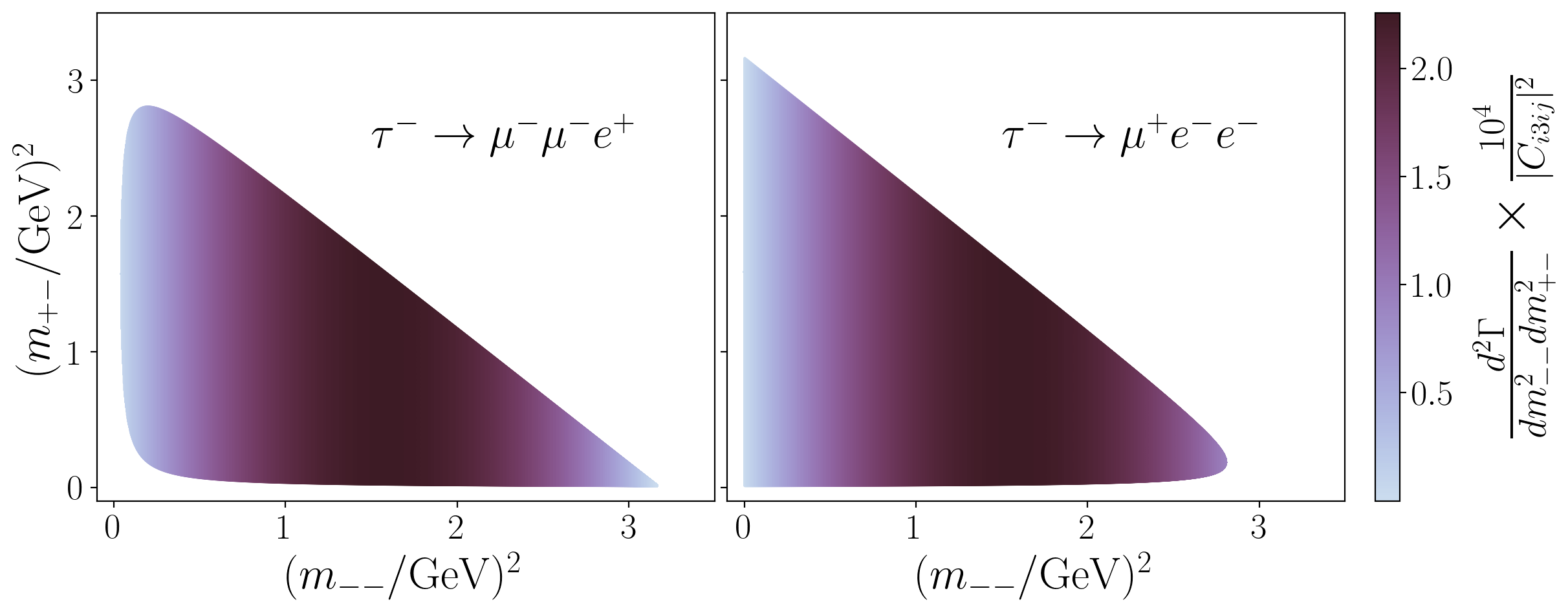}
\includegraphics[width=\textwidth]{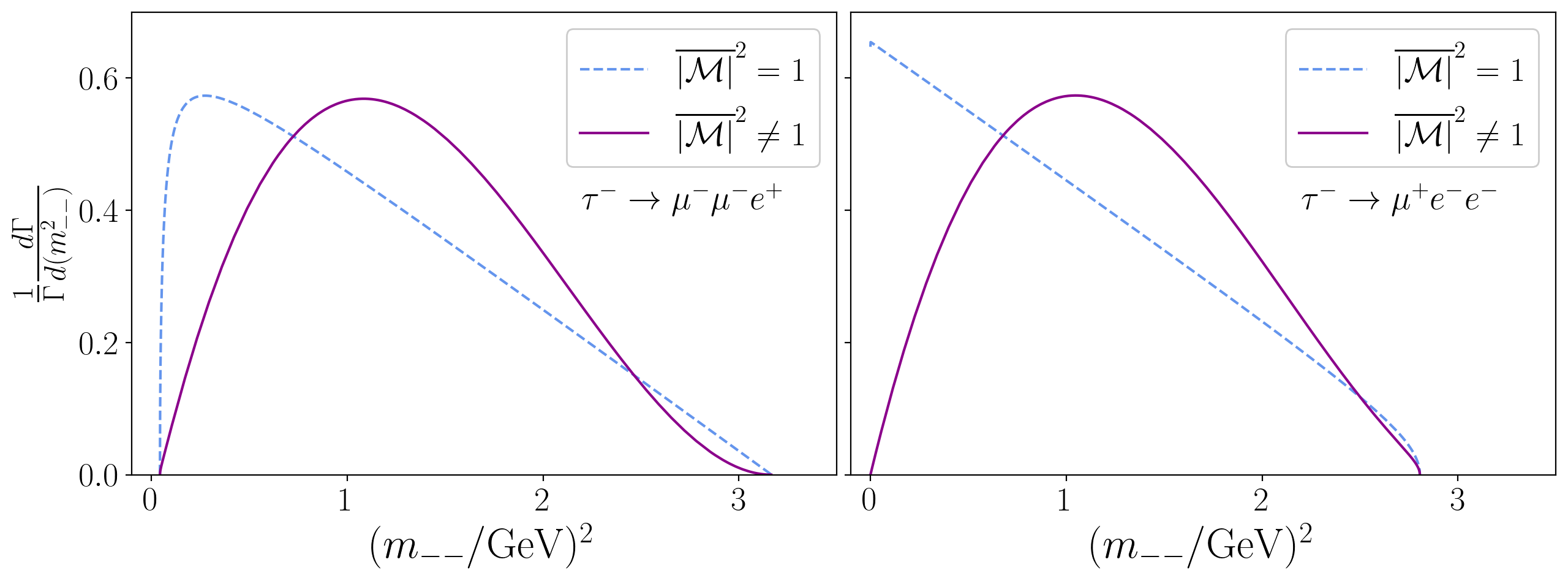}
    \caption{A study of key cLFV tau decay phase space. The top row shows Dalitz plots for the two key decay processes $\tau^- \to \mu^- \mu^- e^+$ (left) and $\tau^- \to \mu^+ e^- e^-$ (right) assuming the dominant contribution is via the left or right-handed vector operator  $C_{i3ij}$, as appears in Eq.~\eqref{eq:phasespace}. The bottom row shows the integrated differential distributions with respect to the observable $m_{--}^2$, illustrating the difference between the kinematic distribution of this model~($\overline{|{\mathcal{M}}|}^2\neq 1$ as per Eq.~\eqref{eq:phasespace}, solid purple line) and the phase space distribution~($\overline{|{\mathcal{M}}|}^2=1$, blue dashed line).}
    \label{fig:Dalitz}
\end{figure}

In the both the electroweak singlet and triplet models, the differential decay rate for $\tau^- \to \ell_i^- {\ell}_i^-\ell_j^+$ is given by
\begin{align}
    \frac{d^2\Gamma (\tau^- \to \ell_i^- {\ell}_i^- \ell_j^+)}{dm_{--}^2\, dm_{+-}^2} &= \frac{1}{256 \pi^3 m_\tau^3} \overline{|\mathcal{M}|}^2 =
    \frac{|C_{i3ij}|^2 }{64 \pi^3 m_\tau^3}  (m_{--}^2- 2m_{\ell_i}^2)(m_\tau^2 +m_{\ell_j}^2 -m_{--}^2 ), \label{eq:phasespace}
\end{align}
where $m_{--}^2= (p_{\ell_i}+p_{\ell'_i})^2$ is the invariant mass of the system of two same-sign leptons in the final state, $m_{+-}^2= (p_{\ell_i}+p_{\ell_j})^2$ is that of two oppositely charged final state leptons, and $\overline{|\mathcal{M}|}^2$ is the spin-averaged matrix element for the process. The squared Wilson coefficient is given by $|C_{i3ij}|^2 =|C^{\rm VRR}_{ee,i3ij}|^2$ in the electroweak singlet models
and by $|C_{i3ij}|^2 =|C^{VLL}_{ee,i3ij}|^2$ in the electroweak triplet models. Thus all models feature the same distributions for the differential decay rates. A so-called \emph{phase space distribution} for the differential decay rate discussed in this section corresponds to setting $\overline{|\mathcal{M}|}^2 =1$.

From Eq.~\eqref{eq:phasespace} there is a flat distribution in $m^2_{+-}$ and a peak in $m^2_{--}$ at $m^2_{--}= \frac{1}{2} (2m_{\ell_i}^2 + m_{\ell_j}+ m_\tau^2)\approx\frac{1}{2} m^2_{\tau}$, where the latter approximation corresponds to neglecting the mass of leptons in the final state. This is illustrated by the Dalitz plots in the top row of Figure~\ref{fig:Dalitz}. The structure of these Dalitz distributions is characteristic of models in which the dominant BSM contribution to these processes is via a RH or LH vector operator, respectively.
Ultimately if these decays are detected with significant multiplicity, an analysis of the three-body phase space could discriminate between different BSM explanations~\cite{Dassinger:2007ru}.

In the bottom row of Figure~\ref{fig:Dalitz} we illustrate the corresponding $\frac{d\Gamma (\tau^- \to \ell_i^- {\ell}_i^- \ell_j^+)}{dm_{--}^2}$ distribution (solid purple), comparing it to a phase space distribution (dashed blue). 
Integration over $m_{+-}^2$ results in a factor
\begin{equation}
m_{+-,\rm max}^2- m_{+-,\rm min}^2 
= \frac{\lambda^{1/2}(m_\tau^2,m_{--}^2,m_{\ell_j}^2) \lambda^{1/2}(m_{--}^2,m_{\ell_i}^2,m_{\ell_i}^2)}{m_{--}} ,
\end{equation}
where $\lambda^{1/2}$ is the square-root of the K\"all\'en function\footnote{$\lambda(x, y, z)\equiv  x^2 + y^2 + z^2 -2xy-2yx-2zx$.}, and the invariant mass $m_{--}^2$ takes values in the range $4m_{\ell_i}^2 <m_{--}^2<(m_\tau-m_{\ell_j})^2$. Integration of the differential decay width results in the expressions shown in Eqs.~(\ref{eq:taubarmuee_k1}) and (\ref{eq:taubaremumu_k2}).

 Even with non-observation in these channels, the expected number of events in different kinematic regions is shown in Figure~\ref{fig:Dalitz} to vary model-dependently (this was also emphasised in Refs.~\cite{Dassinger:2007ru,Goto:2010sn,Celis:2014asa,Bruser:2015yka}, although they did not study these two decay channels). As such, an exclusion should be weighted accordingly to provide the most accurate measure of the present and projected constraint -- although it is not feasible for an experimental collaboration to present constraints on each individual model or even at the EFT level, where interference effects would also need to be considered. Without access to internal Belle/ Belle II information, one cannot easily recast the branching-ratio limits to a specific model, and so we still adopt the values derived using the assumption of a phase space distribution (Table~\ref{tab:LFVtau}). In the absence of events, recasting the limits based on the phase space distribution requires the detection efficiency as a function of the invariant masses $m_{--}^2$ and $m_{+-}^2$.

\section{Neutrino masses}
\label{sec:neutrino}

So far, we have mainly focused on the charged-lepton sector. In this section we turn to neutrino masses and discuss a few possible scenarios how non-zero neutrino masses may be incorporated into these models. 

The most straightforward way to generate non-zero neutrino masses in both the electroweak singlet and triplet models is by introducing three RH sterile neutrinos with similar
$Z_3$ triality charges. The first (second) [third] generation of sterile neutrinos has charge $T= 1$ ($2$) [$3$]. These charges correspond to the transformations
\begin{align}
    \nu_R \to \omega^T \nu_R
\end{align}
where 
$\nu_R$ are the RH neutrinos. Given these assignments, the neutrino Yukawa and mass terms in the Lagrangian are
\begin{align}
   - \mathcal{L} \supset  y_{\nu i} \bar L_i \nu_{Ri} \tilde H
    + \frac12 M_{ij} \overline{(\nu_{Ri})^c}\, \nu_{Rj}
    +\mathrm{h.c.}
    ,
\end{align}
where $i,j=1,2,3$ with repeated indices summed, $\tilde H = i \tau_2 H^*$, $y_{\nu i}$ denotes the neutrino Yukawa couplings, and $M$ is the RH neutrino Majorana mass matrix. The neutrino Dirac mass matrix is diagonal. With exact triality, the RH neutrino Majorana mass matrix is constrained to the form
\begin{align}
    M = \left( \begin{array}{ccc} 0 & M_{12} & 0 \\ M_{12} & 0 & 0 \\ 0 & 0 & M_{33} \end{array} \right).
    \label{eq:initialM}
\end{align}
This is incompatible with the neutrino oscillation data, given that we also have a diagonal neutrino Dirac mass matrix. Therefore lepton triality \emph{must} be broken. This breaking could be achieved by via explicit soft-breaking operators, which then generate the remaining entries of the RH neutrino mass matrix. Alternatively, it can be achieved by introducing a SM singlet complex scalar $S$ with $T=1$ (so that $S\to\omega S$ and $S^* \to \omega^2 S^*$) which leads to the additional triality-preserving Yukawa coupling terms
\begin{align}
    \mathcal{L} &\supset \frac{1}{2} \left[ x_{11} \overline{(\nu_{R1})^c}\, \nu_{R1} + x_{23} \left( \overline{(\nu_{R2})^c}\, \nu_{R3} + \overline{(\nu_{R3})^c}\, \nu_{R2} \right) \right] S \nonumber\\
    &+ \frac{1}{2} \left[ x_{22} \overline{(\nu_{R2})^c}\, \nu_{R2} + x_{13} \left( \overline{(\nu_{R1})^c}\, \nu_{R3} + \overline{(\nu_{R3})^c}\, \nu_{R1}\right) \right] S^* + \mathrm{h.c.}\;.
\end{align}
Triality is then spontaneously broken by a nonzero VEV for $S$, and the zero entries in Eq.~(\ref{eq:initialM}) are now all generated. A diagonal neutrino Dirac mass matrix together with a general RH neutrino Majorana mass matrix is able to accommodate the neutrino oscillation data.

Adopting the type-I seesaw mechanism requires the VEV-generated triality-breaking terms to be of a high scale, not dissimilar to $M_{12}$ and $M_{33}$.\footnote{Given the high spontaneous breaking scale for the discrete $Z_3$ symmetry, the resulting cosmological domain wall problem could be solved by inflating-away the domain walls.} Note that the $S$ and $S^*$ Yukawa terms combine to explicitly break lepton-number conservation. Lepton number is also explicitly broken by a cubic $(a S^3 + \mathrm{h.c.})$ triality-preserving term in the scalar potential, which means that the phase of $S$ is not a Goldstone boson. Neutrino masses can equally well generated using the type-III seesaw mechanism~\cite{Foot:1988aq} with electroweak triplet fermions instead of electroweak singlet fermions.

As was mentioned in the introduction, lepton triality is motivated by discrete flavour symmetries, which break the flavour group to a $Z_3$ subgroup in the charged lepton sector and to a $Z_2$ subgroup in the neutrino sector. The misalignment between the two sectors explains the leptonic mixing matrix with the prime example for this construction being the $A_4$ flavour group. Assuming that the additional BSM physics related to the flavour symmetry is sufficiently decoupled, the $A_4$ flavour symmetry models for neutrino masses mentioned in the introduction  (Refs.~\cite{Altarelli:2005yx,He:2006dk,Ma:2010gs,deAdelhartToorop:2010jxh,deAdelhartToorop:2010nki,Cao:2011df,Holthausen:2012wz,Pascoli:2016wlt,Muramatsu:2016bda}) also yield the phenomenology of charged leptons discussed in this paper.

Aside from triality, the electroweak triplet scalar is that of the type-II seesaw mechanism. A nonzero VEV for $\Delta_T^0$ would therefore contribute some direct Majorana mass terms for the light neutrinos. The VEV is naturally suppressed because the cubic Higgs coupling $H^T i\sigma_2 \mathbf{\Delta}_T^\dagger H$ softly breaks lepton triality. Similarly, the electroweak singlet scalar features in the Zee-Babu model~\cite{Zee:1985id,Babu:1988ki}. See \cite{Nebot:2007bc,Herrero-Garcia:2014hfa,Schmidt:2014zoa} for recent phenomenological studies. 

As it becomes evident from the discussion of the seesaw mechanisms, lepton triality has to be broken to achieve the observed leptonic mixing pattern~\cite{Esteban:2020cvm,nufit}. This can also be seen more generally by considering the Weinberg operator~\cite{Weinberg:1979sa}, 
\begin{equation}
	\mathcal{L}_{5} = -\frac{\kappa_{ij}}{4} (L_i H)^T  C (L_jH) + \mathrm{h.c.} 
		\;,
\end{equation}
where the SU(2) indices are contracted within each pair of parentheses. Lepton triality constrains the Wilson coefficient to take the form\footnote{Note that different neutrino flavours can be singled out using a shift of all triality charges. The maximal 2-3 lepton mixing motivates to shift all triality charges by 2 which results in a Majorana neutrino $\nu_1$ and a Dirac pair $\nu_{2,3}$.}
\begin{equation}
	\kappa = \begin{pmatrix}
		0 & \kappa_{12} & 0\\
		\kappa_{12} & 0 & 0\\
		0 & 0 & \kappa_{33}
	\end{pmatrix}\;.
\end{equation}
Like for the case of the seesaw mechanism, lepton triality has to be broken to explain the observed lepton mixing matrix which may be achieved by introducing a SM singlet complex scalar $S$ with $T=1$ and an effective dimension-6 operators $S^{(*)} (L_i H)^T C (L_jH)$. A similar argument can be made for Dirac neutrinos, because the Yukawa interaction $\bar L_i \nu_{Rj} \tilde H$ and thus the neutrino Dirac mass term contain at most 3 non-zero entries for the lepton triality charges defined in Eq.~\eqref{eq:triality} irrespective of the lepton triality charges of the RH neutrinos.

\section{Conclusions}
\label{sec:conclusions}

Charged lepton flavour-violating decays and other related processes are an important probe of physics beyond the standard model. The discovery of neutrino flavour oscillations implies that such processes must occur at some level, but for them to be observable in practice physics in addition to that responsible for neutrino mass generation must exist. Importantly, significant advances in the search for flavour-violating $\tau$ decays will be made by the Belle II experiment over the next few years, thus opening a new discovery window. In this paper we presented very simple models based on the flavour symmetry structure of lepton triality that feature the decays $\tau^\pm \to \mu^\pm \mu^\pm e^\mp$ and $\tau^\pm \to e^\pm e^\pm \mu^\mp$ as the dominant signals of BSM physics. These decays are driven by the tree-level exchange of doubly-charged scalars shown Fig.~\ref{fig:FeynmanDiagrams}. As illustrated in Figs.~\ref{fig:singletbounds} and \ref{fig:tripletbounds}, the eventual Belle II sensitivity to these processes will see significantly more parameter space explored compared to the present situation, either discovering evidence of new physics or further constraining the possibilities. 

These models are simple examples of minimal SM extensions that single out the $\tau$ sector for the dominant phenomenological signatures. The fact that it proves so easy to do this highlights the importance and relevance of the on-going experimental searches. We expect that our minimal models may be embedded in more complete theories of flavour symmetry and, beyond this, that quite different schemes could also be constructed to achieve a similar purpose.

\section*{Acknowledgements} 
This work was supported in part by Australian Research Council Discovery Project DP200101470 and in part by the Australian Research Council Centre of Excellence for Dark Matter Particle Physics (CDM, CE200100008). It was also supported in part by NSFC (Nos. 12090064, 11975149) and in part by the MOST (Grant No. MOST 106- 2112-M-002- 003-MY3). This manuscript has been authored in part by Fermi Research Alliance, LLC under Contract No. DE-AC02-07CH11359 with the U.S. Department of Energy, Office of Science, Office of High Energy Physics.

\appendix

\section{SMEFT}
\label{sec:SMEFT}
In addition to the renormalisable part of the Lagrangian we introduce dimension-6 SMEFT operators. We are particularly interested in operators which violate lepton flavour and are consistent with the $Z_3$ symmetry. The only relevant operators in the Warsaw basis~\cite{Grzadkowski:2010es} are the 4-lepton operators
\begin{align}
\mathcal{L}_{6}  & =C^{ll} (\bar L \gamma_\mu L )(\bar L \gamma^\mu L) 
+ C^{ee} (\bar e_R \gamma_\mu e_R)(\bar e_R \gamma^\mu e_R)
+C^{le} (\bar L\gamma_\mu L )(\bar e_R \gamma^\mu e_R),
\end{align}
where we do not explicitly specify the flavour indices. In case flavour indices are important, they are written as subscripts in the order of the fermion flavours in the operators, e.g.~$C^{ll}_{abcd}$ is the Wilson coefficient of operator $(\bar L_a \gamma_\mu L_b)(\bar L_c \gamma^\mu L_d)$.

The electroweak singlet models lead to
\begin{align}
C^{ee}_{3322} & =\frac{|f_1|^2}{2m_{k_1}^2} &
C^{ee}_{1111} & = \frac{|f_2|^2}{8m_{k_1}^2} &
C^{ee}_{1312} & = C^{ee*}_{3121} = \frac{f_1f_2^*}{4m_{k_1}^2}\\
C^{ee}_{3311} & =\frac{|g_1|^2}{2m_{k_2}^2} &
C^{ee}_{2222} & = \frac{|g_2|^2}{8m_{k_2}^2} &
C^{ee}_{2321} & = C^{ee*}_{3212} = \frac{g_1g_2^*}{4m_{k_2}^2}
,
\end{align}
where we removed equivalent Wilson coefficients and only keep one of the equivalent flavour combinations. For the electroweak triplet models we find
\begin{align}
C^{ll}_{3322} & = C^{ll}_{3223} = \frac{|f_1|^2}{4m_{\Delta_1}^2} &
C^{ll}_{1111} & = \frac{|f_2|^2}{8 m_{\Delta_1}^2} &
C^{ll}_{1312} & = C^{ll*}_{3121} = \frac{f_1f_2^*}{4m_{\Delta_1}^2}\\
C^{ll}_{3311} & = C^{ll}_{3113} = \frac{|g_1|^2}{4m_{\Delta_2}^2} &
C^{ll}_{2222} & = \frac{|g_2|^2}{8 m_{\Delta_2}^2} &
C^{ll}_{2321} & = C^{ll*}_{3212} = \frac{g_1g_2^*}{4m_{\Delta_2}^2}
.
\end{align}
As lepton triality protects the flavour structure of the operators, no operator violating lepton triality is generated by renormalisation group corrections. Thus in this analysis we neglect renormalisation group corrections.

\section{LEFT}
\label{sec:LEFT}
As the $Z_3$ symmetry only allows lepton-flavour-violating 4-fermion operators at dimension-6 in SMEFT, we consider only leptonic 4-fermion interactions in LEFT~\cite{Jenkins:2017jig}:
\begin{align}
    \mathcal{L} 
    & = 
    C^{VLL}_{ee} (\bar e \gamma^\mu e_L)(\bar e \gamma_\mu e_L)
    +C^{VRR}_{ee} (\bar e \gamma^\mu e_R)(\bar e \gamma_\mu e_R)
    +C^{VLR}_{ee} (\bar e \gamma^\mu e_L)(\bar e \gamma_\mu e_R)
    \\\nonumber
&
    +C^{VLL}_{\nu e} (\bar \nu \gamma^\mu \nu_L)(\bar e \gamma_\mu e_L)
    +C^{VLR}_{\nu e} (\bar \nu \gamma^\mu \nu_L)(\bar e \gamma_\mu e_R).
\end{align}
Similarly to the SMEFT operators, we do not explicitly specify the flavour indices, unless needed. In case flavour indices are important, they are written as subscripts in the order of the fermion flavours in the operator, e.g.~$C^{VLL}_{ee,abcd}$ is the Wilson coefficient of $(\bar e_a \gamma^\mu P_L e_b)(\bar e_c \gamma_\mu P_L e_d)$.
We do not include the 4-neutrino operator, because it is not relevant for the discussion of the phenomenology. Other lepton-flavour-violating Wilson coefficients are strongly suppressed by the unitarity of the PMNS mixing matrix and are neglected.

The matching to LEFT operators is given by~\cite{Jenkins:2017jig}
\begin{align}
C^{VLL}_{ee,ijkl} & = C^{ll}_{ijkl} - \frac{g_Z^2}{4m_Z^2} [ Z_{e_L}]_{ij} [ Z_{e_L}]_{kl} - \frac{g_Z^2}{4m_Z^2} [Z_{e_L}]_{il} [Z_{e_L}]_{kj},
\\
C^{VRR}_{ee,ijkl} & = C^{ee}_{ijkl} - \frac{g_Z^2}{4m_Z^2} [ Z_{e_R}]_{ij} [ Z_{e_R}]_{kl} - \frac{g_Z^2}{4m_Z^2} [Z_{e_R}]_{il} [Z_{e_R}]_{kj},
\\
C^{VLR}_{ee,ijkl} & = C^{le}_{ijkl} - \frac{g_Z^2}{m_Z^2} [ Z_{e_R}]_{ij} [ Z_{e_R}]_{kl} ,
\\\label{eq:CVLLnue}
    C^{VLL}_{\nu e ,ijkl} & = C^{ll}_{ijkl} + C^{ll}_{klij} -\frac{g^2}{2 m_W^2} [W_l]_{il} [W_l]_{jk}^* -\frac{g_Z^2}{m_Z^2} [Z_\nu]_{ij} [Z_{e_L}]_{kl},
    \\\label{eq:CVLRnue}
    C^{VLR}_{\nu e, ijkl} &=C^{le}_{ijkl} - \frac{g_Z^2}{m_Z^2} [ Z_\nu]_{ij} [ Z_{e_R}]_{kl},
\end{align}
with the SM contributions via $W$ and $Z$ boson exchange 
\begin{align}
    [Z_\nu]_{pr} & = \frac12 \delta_{pr}, 
    & 
    [Z_{e_L}]_{pr} & = \left(-\frac12 + s_w^2\right) \delta_{pr},
    &
    [Z_{e_R}]_{pr} & = s_w^2 \delta_{pr},
    &
    [W_l]_{pr} & = \delta_{pr}.
\end{align}
The Fermi constant in the SM is given by $2\sqrt{2}G_F=\frac{g^2}{2m_W^2} = \frac{g_Z^2}{2m_Z^2}$ and thus Eqs.~(\ref{eq:CVLLnue},\ref{eq:CVLRnue}) result in the SM contribution shown in Eq.~\eqref{eq:nu-e-SM}.
Renormalisation group corrections are dominated by QED running and thus generally small, so we neglect them throughout this analysis.

\bibliography{refs}

\end{document}